\documentclass[%
 reprint,
 amsmath,amssymb,
 aps,
]{revtex4-2}

\usepackage{amssymb,amsmath,amsfonts,mathtools,mathrsfs} 
\usepackage{mdframed}
\usepackage{color}
\usepackage{graphicx}
\usepackage{dcolumn}
\usepackage{bm}
\usepackage{bbm}
\usepackage{hyperref}
\hypersetup{
    colorlinks=false,
    linkcolor=blue,
    filecolor=magenta,      
    urlcolor=cyan,
    }

\newcommand\fa[1]{{
#1}}

\begin{document}

\preprint{APS/123-QED}

\title{Multipoint fishnet Feynman diagrams: sequential splitting}

\author{Francesco Aprile$^{a}$ and\, Enrico Olivucci$^{b}$\\ ~}

\affiliation{%
\\
\centerline{\it \({}^{a}\)
Departamento de F\'isica Te\'orica \& IPARCOS, Facultad de
Ciencias F\'isicas,}
Universidad Complutense de Madrid,
28040 Madrid, Spain
 \\
\({}^{b}\) 
Perimeter Institute for Theoretical Physics, Waterloo, Ontario N2L 2Y5, Canada
\\
}

\begin{abstract}
We study fishnet Feynman diagrams 
defined by a certain triangulation of a planar $n$-gon,
with \fa{massless} scalars  propagating along and across the cuts.
Our solution theory uses the technique of Separation of Variables, 
in combination with the theory of symmetric polynomials and Mellin 
space.  The $n$-point split-ladders are solved by a recursion 
where all building blocks are made fully explicit.  In particular, we find an 
elegant formula for  the coefficient functions of the light-cone leading logs.
When the diagram grows into a fishnet, we obtain new results 
exploiting a Cauchy identity decomposition of the measure over separated variables.
This leads to an elementary proof of the Basso-Dixon formula at $4$-points,
while at $n$-points it provides a natural OPE-like stratification of the diagram. 
Finally, we propose an independent approach based on `stampede' combinatorics to study the light-cone behaviour 
of the diagrams as the partition function of a certain vertex model.
\end{abstract}

\maketitle

\begin{figure}[b]
\includegraphics[scale=0.48]{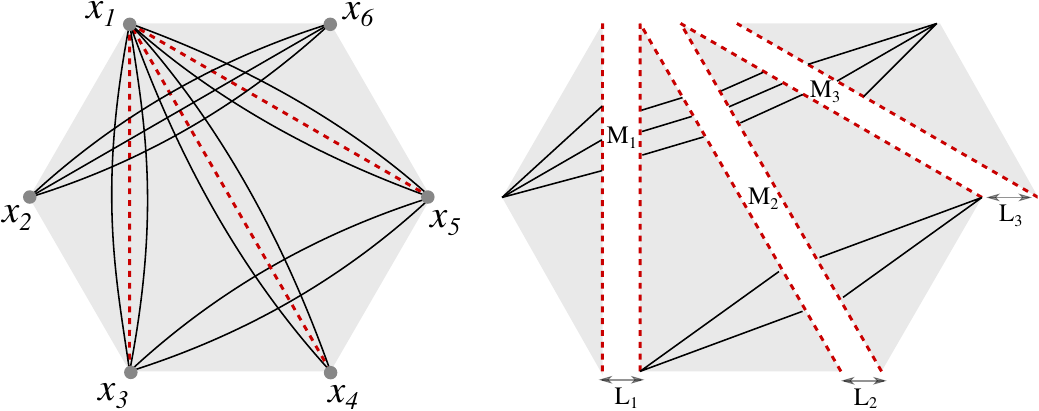}
\caption{A 6pt diagram on the plane and its SoV triangulation. 
There are $L_{i=1,2,3}=2$ propagators along, and $M_{i=1,3}=3$, $M_2=5$, 
propagators transverse to the cuts, drawn in red.}
\label{triang_ratios}
\end{figure}

\begin{figure*}[t]
\centering
\includegraphics[width=0.98\textwidth]{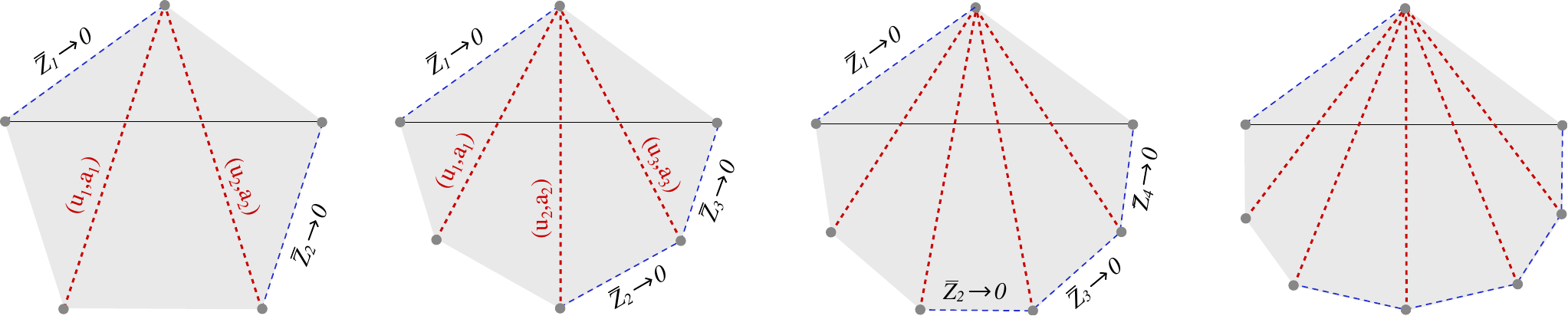}
\caption{Split-Ladder Feynman diagram in the sequential triangulation. It follows 
from \eqref{crossratios} that the limit $\bar Z_k \to 0$ can be realised by sending the 
dashed blue edges on the light-cone. This kinematical region of the Fishnet 
corresponds to a choice of signature $(-,-,\dots, -)$ in the integration over SoV rapidities.}
\label{slicing}
\vspace{-3 mm}
\end{figure*}

{\bf Introduction.}
Feynman diagrams \cite{Feynman:1949hz} are  central for our understanding 
of Nature at the quantum level. The case of collider physics is a perfect example 
of this paradigma. However, diagrams are often hard to compute and alternative 
routes are needed in the study of a relativistic quantum process.
In this letter we embrace this point of view and use an integrability-based 
technique, known as Separation of Variables (SoV),
to study a class of \emph{multipoint} Feynman diagrams in $4d$,
whose complexity depends on a \emph{fishnet} of \fa{massless} scalar propagators.
Our diagrams  lie on a \emph{planar} triangulation of an $n$-gon, and
their SoV representation can be obtained by gluing together objects 
assigned to triangles. The general pattern is presented in \cite{Olivucci:2023tnw}.
Here we will focus on splitting a $4$pt triangulation sequentially, so 
to generate a particular class of Feynman diagrams. See~FIG.~\ref{triang_ratios}\,-\,\ref{slicing}. 

While the SoV representation is not new \cite{Maillet:2019nsy, Basso:2018cvy, Derkachov:2019tzo, Derkachov:2020zvv},  
some practical challenges 
with it remained unexplored until now, especially for $n\ge 5$. 
First, each triangle carries an $R$-matrix tensor structure
contribution from the underlying spin-chain formalism. 
Second, for higher-pts there are many possible configurations of propagators.
Third, there is a variety of kinematical limits $x_{ij}^2\rightarrow 0$ that one 
might want to consider.
In this situation, what we expect from SoV is the ability to make manifest new structures,
otherwise hidden in the Feynman representation, and  the ability to provide 
data in a simple way, e.g. by expansions around light-cone limits. 
At $4$pt all of this is well established. Indeed, the SoV representation  
allowed~\cite{Derkachov:2019tzo} to prove the conjecture 
of~\cite{Basso:2017jwq} about rectangular fishnets being 
equal to determinants of ladder integrals.

In this letter we set the stage for investigating emergent structures in
multi-point fishnet diagrams that lie on the plane. 
First,  we shall understand the simplest diagrams, i.e.~split-ladders.
Second, we will massage the SoV representation so to make 
it as explicit as possible. For the first task, we will show that
the integration of $n$-pt split ladders reduces to a recursion, 
and that the relevant $R$-matrix construction admits a simple  
closed-form expression. 
Remarkably, the coefficient functions of the light-cone leading logs
take the elegant form
\begin{equation}
\label{maxlogs_1}
\!\!\!\begin{array}{rl}
\displaystyle \frac{\tt MPLs}{(W_{n-3}\!-\!\bar{W}_{n-3})}\!=&
\!\!\!\!\!\!{\displaystyle \sum_{\substack{a_2\in\mathbb{N}\\ ...\\a_{n-3}\in\mathbb{N}}}
\!\!\prod_{i\ge 2}\frac{ \mathcal{D}^{(a_i)}_{i,i-1}}{a_{i}^{L_{i} }}}
\!\! \left[ {\displaystyle \frac{ {\rm Li}_{L_1}(W_1) - {\rm Li}_{L_1}(\bar{W}_1) }{W_1-\bar{W}_1} }\right]
\end{array}
\vspace{-1mm}
\end{equation}
where at any $n\ge5$ the differential operator is
\begin{equation}
\label{intro_diff}
\mathcal{D}^{(a_i)}_{i,i-1}=\frac{(-\bar{W}_{i} \partial_{\bar{W}_{i-1}} - W_{i} \partial_{W_{i-1}})^{a_i-1}}{(a_i-1)!}\,.
\end{equation}

For the second task, we will present a 
Cauchy identity decomposition of the SoV measure that makes 
transparent the derivation of the $4$pt determinant of 
\cite{Basso:2017jwq, Basso:2021omx}, and paves the 
way to the understanding of the $n$-pt split-fishnet diagrams.
Finally, we make an intriguing observation inspired by 
stampede methods \cite{Olivucci:2021pss} that relates 
the coefficient functions of the light-cone leading logs with the partition 
function of some vertex model.\\

{\bf SoV diagrammatic rules.}
Draw an $n$-gon \emph{on the plane} and a triangulation of it into $n-2$ triangles. 
Edges between non-adjacent external points cut the diagram, and are $n-3$.
To each cut assign an integer $L_j$, that counts propagators along the cut, 
and another integer $M_j$,  that counts propagators across the cut. 
For illustration see FIG.~\ref{triang_ratios}.

\begin{figure}[t]
\includegraphics[scale=0.55]{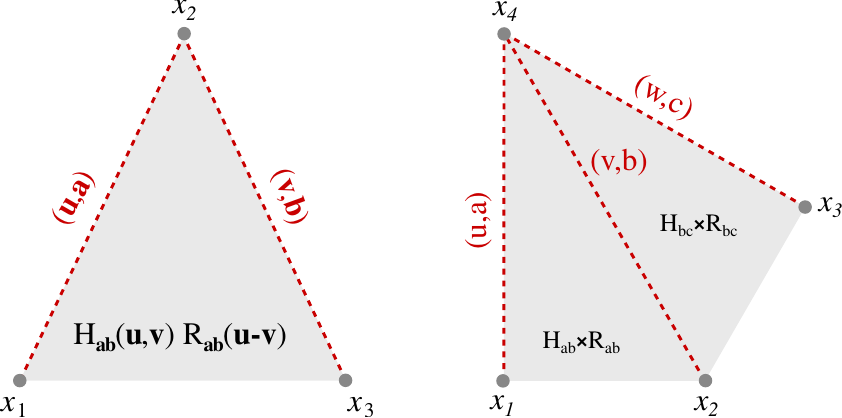}
\caption{\textbf{Left:} A tile of the SoV triangulation of a Fishnet integral carrying 
excitations $(\mathbf{u},\mathbf a)$ and $(\mathbf{v},\mathbf b)$ on its cuts 
(red dashed edges). \textbf{Right:} The quadrilateral obtained by gluing two 
triangles along a common edge.}
\label{tile}
\vspace{-4 mm}
\end{figure}

The SoV representation of such diagrams is a certain integral that can be read from 
the following set of rules. The $j$-th cut carries rapidities ${\bf u}_j=(u_{j_1},\ldots, u_{j_{M_j}})$  
and quantum numbers ${\bf a}_j=(a_{j_1},\ldots, a_{j_{M_j}})$. Here $u\in\mathbb{R}$,
while $a\in \mathbb{N}$ labels an $su(2)$ rep $V_a$ of spin  $\frac{a-1}{2}$. 
Each cut contributes to the integrand with a measure $\mu$, and an energy factor $E$. 
A triangle filling in between two cuts, represented in FIG.~\ref{tile}, contributes 
with an interaction that couples the variables on the two cuts. This interaction 
is factorised into a scalar part  $H_{ab}(u,v)$, and matrix part $R_{ab}({\bf i}u-{\bf i}v)$ 
with ${\bf i}=\sqrt{-1}$. 
Bold font notation wants to abbreviate multivariable contributions, which we will 
make explicit in \eqref{boldfont_notation}-\eqref{boldRmain}. Note that 
$R_{ab}\in{\rm End}( V_a\otimes V_b)$,  in particular, $R_{ab}:=(R_{ab})_{ij,kl}$.
Note also that the triangles are oriented, indeed $H_{ab}(u,v)\neq H_{ba}(v,u)$.

To read off the spacetime dependence from the triangulation we introduce 
a pair of cross ratios for any quadrilateral obtained by gluing two triangles that share a cut.
With reference to FIG.\ref{tile}, we define the ratios as
\begin{equation}
\label{crossratios}
Z \bar Z = \frac{ x_{14}^2x_{23}^2}{x_{12}^2 x_{34}^2}\qquad;\qquad 
(1-Z)(1-\bar Z) = \frac{ x_{13}^2 x_{24}^2}{x_{12}^2 x_{34}^2} 
\end{equation}
Any such quadrilateral contributes to the SoV integrand with the factor
\begin{equation*}
 \otimes_{k=1}^{M}\left[ \rho^{{\bf i}u_k-\frac{1}{2}} \times e^{{\bf i} \theta {J}_{3, a_{k}}}\right]
 \qquad;\qquad\begin{array}{c} \rho\equiv Z \bar Z\\[.2cm] e^{2 {\bf i} \theta}=Z/\bar Z \,.
 \end{array}\vspace{0.5 mm}
\end{equation*}
Here ${J}_{3,a} \in \text{End}(V_{a})$ is the generator of rotations 
on the plane of the $n$-gon.  
There is a scalar part, $\rho^{\sum ({\bf i}u_k-\frac{1}{2})}$, and matrix part 
that we denote by ${\cal T}_{\underline{\bf a}}$.
Altogether the SoV integrand for a diagram with propagators 
$\underline{L}=L_1\ldots L_{n-3}$ and $\underline{M}=M_1\ldots M_{n-3}$ is given by
\begin{widetext}
\begin{equation}
 \label{calFdef}
\mathcal{F}_{\underline{L},\underline{M}}= 
\ \prod_{j=1}^{n-3}{ \mu}_{ {\bf a}_j}({\bf u}_j) \times E_{\mathbf{a}_j}(\mathbf{u}_j)^{L_j+M_j}\times\!
\left(\rho_j\right)^{\sum_{k=1}^{M_j} \left({\bf i} u_{jk} -\frac{1}{2}\right)} \times
 \prod_{j=2}^{n-3}{\bf H}_{\mathbf{a}_{j-1},\mathbf{a}_{j}}({\bf u}_{j-1},{\bf u}_{j}) \times 
 {\cal T}_{ \underline{\bf a}}\left[ \theta_1,\ldots; {\bf u}_1,\ldots \right] \vspace{-1 mm}
\end{equation}
\end{widetext}
where $\underline{\bf a}=({\bf a}_1,\ldots,{\bf a}_{n-3})$ and we used\vspace{-2 mm}
\begin{align}
\begin{aligned}
\label{boldfont_notation}
{\bf H}_{\mathbf{a},\mathbf{b}}({\bf u},{\bf v})=&\prod_{k=1}^{M} \prod_{h=1}^{N}H_{a_{h},b_{k}}(u_{h},v_{k}) \\
E_{\mathbf a}(\mathbf u) =& \prod_{m=1}^M E_{a_m}(u_m) = \prod_{m=1}^M \frac{1}{\left(u_m^2 +\frac{a_m^2}{4}\right)}
\end{aligned}
\end{align}
The functions $\mu$ and $H_{ab}$ are taken from \cite{Basso:2017jwq,Basso:2018cvy,Derkachov:2019tzo,Olivucci:2023tnw}
and reported in the suppl.~mat.~of this letter. 
The contribution ${\cal T}_{\underline{\bf a}}$ in \eqref{calFdef} 
is the trace on a product of $R$-matrices \vspace{-2 mm}
\begin{equation}
\label{boldRmain}
 \,{\bf R}_{\mathbf a\mathbf b}(\mathbf {u}-\mathbf{v}) = \prod_{k=1}^{M} \prod_{h=1}^{N} {R}_{a_kb_{N-h+1}}({\bf i}u_k-{\bf i}v_{N-h+1})\,,
\end{equation}\vspace{-2 mm}
\begin{equation}
\label{T_parition}
\! {\cal T}_{\underline{\mathbf{a}}}=\text{Tr}_{\underline{\mathbf a}} \left[
\displaystyle{ \bigotimes _{\substack{  \\1\leq j\leq n-3\\[.05cm] \,1\leq m\leq M_j}}}\!\!\!\left(\frac{Z_j}{\bar{Z}_j}\right)^{\!\!{J}_{3;a_{jm} }}\!\!\!\!\!\cdot \ \ 
\prod_{l=2}^{n-3} \! {\bf R}_{\mathbf{a}_{l-1},\mathbf{a}_{l}}({\bf i} \mathbf{u}_{l-1}\! -\! {\bf i} \mathbf{u}_{l}) \right]\notag
\end{equation}
where $\text{Tr}_{\underline{\mathbf a}} \equiv\text{Tr}_{\mathbf{a}_1}\!\!\cdots  
\text{Tr}_{\mathbf{a}_{n-3}}$ is running over the indices of all the spaces $V_{a_{j_m}}$.
The matrix ${R}_{a_hb_k}$ acts as the identity in all other spaces 
which are not $V_{a_{k}}$ and $V_{b_{h}}$. For example, 
the expression ${R}_{ab}{R}_{ac}$ contains a {matrix} product only in the space $V_a$.
Finally, \eqref{calFdef} is integrated over rapidities and 
summed over quantum numbers, i.e.~
\begin{equation}
\label{muAU}
\mathcal{I}_{\underline{L},\underline{M}}= 
\sum_{a_1=1}^\infty \cdots  \sum_{a_M=1}^\infty \int d {u_1} \cdots \int d{u_M}\, \mathcal{F}_{\underline{L},\underline{M}}
\vspace{-2mm}
\end{equation}
The $n$-point split-ladders are obtained from \eqref{calFdef} by 
setting $L_i\ge1$ with all the $M_i=1$.   
From these, we can think of growing a fishnet of propagators
by increasing the $M_i>1$. The $4$-pt representative is 
$\mathcal{F}_{L_1,M_1}$. As soon as $n\ge 5$, no matter the values 
of $L_i,M_i$, the matrix part ${\cal T}_{\underline{\bf a}}$ becomes non trivial.\\

\vspace{-1 mm}

{\bf Strategies and signatures.}
$\mathcal{F}_{\underline{L},\underline{M}}$ 
is a meromorphic function of the  $u_j$ .
Hence, the integration on the real axis in \eqref{muAU} can be performed 
via Cauchy's theorem by closing the contour with an arc at $\infty$ for each $u_j$.
The contour is closed in the lower half or the upper half complex plane
depending on the kinematical \emph{regions}, $\rho_j<1$ or $\rho_j>1$.
We encode this in a \emph{signature} $\sigma_j=\pm$.
Performing the residue integration accordingly, 
the result is an expansion of the diagram
with the form of a log-stratification
\vspace{-2mm}
\begin{equation}\label{log_strata}
\mathcal{I}_{\underline{L},\underline{M}}=
\sum_{ \{k_i\ge 0\}} 
\mathbb{F}^{\{\sigma_i\}}_{\underline{k}}(\{Z_i,\bar{Z}_i\}) \times \prod_{j=1}^{n-3} \log^{k_i}\!\rho_j^{ -\sigma_j }\,,
\vspace{-2mm}
\end{equation}
From SoV the task is to obtain all $\mathbb{F}$ coefficient functions. 
The ``maxlog", defined as the 
$\mathbb{F}$ for which $\sum k_i$ is max, is the same as the light-cone leading 
log, and therefore plays a special role.
Now, since for $n\ge 5$ there are a growing number of signatures,
our strategy will be to study a particular one, and perform the integration.
Our choice of signature in this letter is $(-,\dots,-)$, and corresponds to the small-$\rho_j$ expansion.
We picked this particular one because it gives rise to a nice recursion w.r.t.~the number of external points.\\
\vspace{-1mm}

{\bf Split-Ladders as Mellin-Barnes integrals.}
The $n$pt split ladders have by definition $M_i=1$ in \eqref{calFdef},
and in particular the SoV measure becomes abelian.
We shall denote them by 
$\mathcal{L}_{\underline{L}}:={\cal F}_{\underline{L},1\ldots,1}$.
Since the measure $\mu$ is abelian, we will be able to apply standard Mellin-Barnes 
techniques (MB).  
Within the $(-,-\dots-)$ signature is convenient to change variables from $u_j$ to $s_j$, as follows,
\begin{equation}\label{chvar_t}
{\bf i}\, u_j= \tfrac{a_j}{2}+s_j\,,
\end{equation}
so that the integrand is `canonical' as a MB integrand:
\begin{align}
\label{nsplitladder}
&
\mathcal{L}_{L_1,\ldots L_{n-3}}= \\
&
\!\!\frac{\rho_1^{s_1}}{\Gamma[1-s_1](-s_1)^{L_1}}\!\Bigg[ \prod_{j=2}^{n-3}\frac{  \rho_i^{s_j} \Gamma[s_{j}-s_{j-1}] }{(-s_j)^{L_j}}\Bigg] \!\frac{\!\!\Gamma[-s_{n-3}]\ \mathcal{T}_{\underline{a} }}{\prod_{i=1}^{n-3} (a_i+s_i)^{L_i}}, \notag
\end{align}
where $\underline{a}=(a_1,\ldots,a_{n-3})$. 
The matrix part reads
\begin{equation*}
{\cal T}_{ \underline{a} }=
\frac{{\cal P}_{\underline{a}}(Z_1,\bar{Z}_1,\ldots; s_1\ldots)\,\prod_j a_j}{\ \prod_{i=2}^{n-3}\Gamma[1+a_{j-1}+s_{j-1}-s_{j}]}  \frac{\Gamma[a_{1}+s_{1}]}{\Gamma[1+a_{n-3}+s_{n-3}]}.
\end{equation*}
and it contains an interesting polynomial ${\cal P}_{\underline{a}}$, that we study later in \eqref{the_polynomial}.
For the moment note that the SoV representation \emph{sequentiates}, 
see also FIG.~\ref{slicing}. Thus the natural way to integrate it is by iteration, 
starting from $s_{n-3}$, $s_{n-2}$, etc.~until reaching $s_1$. 
By doing so we implicitly assume $\rho_{n-3}<1$ first, 
then $\rho_{n-2}< 1$, etc.
Each time we perform an integration, we effectively \emph{eat} a 
triangular slice from the $n$-gon~as in FIG.\ref{slicing}. Henceforth, 
we produce  a set of reduced integrands whose structure is 
that of a split-ladder with one less external point. 
This structure repeats and a recursion relation emerges.\\
\vspace{-1mm}

{\bf Maxlog function and R-matrix at the pole.}  
The maxlogs originate from higher degree poles in $s_j=0$, 
upon integration by parts ({\tt ibp}). Namely, if $X$ and $\ell$ are generic,
the following is true:
\begin{equation*}
\int\!\!\frac{X}{(-s)^{\ell+1}}=-\int\!\frac{X}{\ell!}\,\partial_s^{\ell} \bigg(\frac{1}{s}\bigg)\ \underbracket{\ = \ }_{{\tt ibp}}\ \int\!\frac{ (-)^{\ell+1}}{\ell!}\frac{\partial_s^\ell X}{s} \,.
\end{equation*} 
A particular maxlog is found when  
all derivatives hit sequentially the exponent of $\rho_j$'s, 
producing the term $\prod_j (\log\rho_{j})^{L_j}$. 
This is not the only one we expect: when two base-points of a triangle 
effectively approach each other, see FIG.\ref{slicing}, 
propagators lying on different adjacent cuts will pile up, and new 
contributions are generated. 
 These new contributions come from {\tt ibp}
 derivatives in $s_j$ hitting the $\Gamma[s_j-s_{j-1}]$. As a result,
the order of the \emph{next}  pole, $s_{j-1}=0$, increases. 
By taking this into account, 
\begin{equation*}
{\cal I}_{\underline{L}}\Big|_{\textnormal{maxlog}}=
\sum_{\substack{ 0\leq i_{n-3}\leq L_{n-3}\\...  \\[.1cm] 0\leq i_2\leq L_2+j_{(3)} }}
\!\!\!\mathbb{F}^{-...-}_{i_1,\dots,i_{n-3}}(\{Z_i,\bar{Z}_i\}) \
\prod_{k\ge 1} \log^{i_k}\!\rho_{k}  \,,
\end{equation*}
where $i_1 =  L_1+j_{ (1) }$ and $ j_{(k)}=\ \textstyle{ \sum_{m> k} (L_m -i_m)}$.
%
%
An explicit computation shows that maxlogs are an 
evaluation formula  at $s_j=0$ that we can write as
\begin{align}\label{Tas_string}
\mathbb{F}^{-...-}_{i_1,\dots,i_{n-3}}&=
 \sum_{\underline{a}}  \frac{ {\cal T}_{ \underline{a} }(0) }{ \prod_{k\ge 1} a_k^{L_k} }\times
\prod_{k\ge 1}\frac{ (-)^{L_k{+1}}}{i_k!}\,,\\[.2cm]
\mathcal{T}_{\underline{a}}({0})&=
\mathcal{D}^{(a_{n-3})}_{n-3,n-2}\ldots \mathcal{D}^{(a_2)}_{\,2,1}\left[ \frac{W_1^{a_1}-\bar{W}_1^{a_1}}{W_1-\bar{W}_1}\right]\,,
\end{align}
in term of the operators \eqref{intro_diff},
where $W_1=Z_1$, $\bar W_1=\bar Z_1$ and $W_{i\ge 2}={Z_i}W_{i-1}$, similarly for $\bar W_{i\ge 2}$. 
Nicely enough, the sum over $a_1\in\mathbb{N}$ in \eqref{Tas_string} can be done, and yields the rhs of formula \eqref{maxlogs_1}.
Note that the split-ladders  are pure transcendental function of $(Z_i,\bar Z_i)$, multiplied 
by a rational prefactor, i.e.~the leading discontinuity. From the maxlog we find that the latter is $(W_{n-3}-\bar{W}_{n-3})^{-1}=(\prod Z_i -\prod \bar{Z}_i)^{-1}$.
This fact has been checked independently from the 
Feynman representation of the diagram, 
see eg.~\cite{Cachazo:2008vp,Drummond:2013nda}.\\
\vspace{-1mm}

{\bf Twisted traces and  polynomials.} 
The SoV representation provides concrete formulae for any 
coefficient function in \eqref{log_strata}, not just the maxlogs.
The only complication comes from the $R$-matrix and the polynomial
\begin{align}
\label{the_polynomial}
\!\!\!\!{\cal P}_{\underline{a}}=& \ \ \rho_1^{\frac{a_1-1}{2}}  \prod_{i=2}^{n-3} (-1)^{a_i-1} (\rho_i)^{\frac{a_i-1}{2}} \times \\[-.2cm]
&\ \ \ \ \ 
\ \ \ \times 
\text{Tr}_{\underline{a}} \left[
\displaystyle{ \bigotimes _{\substack{  \\1\leq j\leq n-3}}}\!\!\!\left(\frac{Z_j}{\bar{Z}_j}\right)^{\!\!{J}_{3;a_{j} }}\!\!\!\!\!\cdot \ \ 
\prod_{l=2}^{n-3} \! {R}_{{a}_{l-1},{a}_{l}} \right]\notag\,.
\end{align}
Nicely enough, we have found an explicit formula for this trace. 
Let us introduce first,
\begin{align}\label{little_r_formula}
&r_{ai;bj}(U)=\sum_{k\ge0}
 (1+j-a+i-U)_{b-1-j-k}\,(a-i)_k \times \notag \\
 &\!\!\times\frac{(k-U-i)_{j-k} (b-j-k)_k (1+i-k)_k(1+j-k)_k}{(k!)^2}. \notag
\end{align}
%
%
Then, the simplest 5pt case,  ${\cal P}_{ab}$, is given by 
\begin{align}
{\cal P}_{ab}=\sum_{i=0}^{a-1}\sum_{j=0}^{b-1}  r_{ai,bj}(U_{12})\  Z_1^{i} Z_2^{j} \bar{Z}_1^{a-1-i} \bar{Z}_2^{b-1-j}\,,
\end{align}
with $U_{12}\equiv s_1-s_2$. Here the sum over  
$Z^{\#_1}\bar{Z}^{\#_2}$ runs over the eigenvalues of ${J}_{3;a},\,{J}_{3;b}$.
The generalisation to higher points is straightforward,
\begin{equation}\label{polyfinal}
\!{\cal P}_{\underline{a}}=\!\!\!\!\!\!\!\!
\sum_{\substack{ \!\! j_1 \le a_1-1 \\ ...\\[1mm] \!\! j_{n-3}\le a_{n-3}-1 } }\!\!
\! \prod_{k=2}^{n-3}  r_{a_{k-1}j_{k-1},a_k j_k}(U_{k-1,k}) Z_1^{j_1}\!...\,\bar{Z}_1^{a_1\!-1\!-j_1}...\Big],\notag
\end{equation}
and works upon checking on (many!) explicit examples. 
\noindent
At the special point $U=0$ we found
\begin{equation*}
(R_{ab})_{ij,kl}(U=0)= 
\Gamma[b]\, \Big[\,^{i}_{\rule{0pt}{.25cm}j}\Big]\Big[\,^{a-1-i}_{\rule{0pt}{.25cm}b-1-j}\Big]
\delta_{i+j,k+l} \,,
 \label{r_atpole}
\end{equation*}
hence the trace, ie.~$(-)^{b-1}r_{ai;bj}(0)$, simplifies drastically.\\
\vspace{-1mm}

{\bf Cauchy identity tool for multipoint fishnets.}
The crucial novelty for fishnets, wrt split-ladders, is the 
non abelian measure. To deal with it systematically, 
we use the dual Cauchy identity for Schur 
polynomials $P_{\underline{\lambda}}$, see e.g.~\cite{Macdonald_book},
and rewrite the measure as
\begin{align}
\mu_{\bf a}({\bf u})
&=\frac{ {\rm VdM}({\bf s}){\rm VdM}({\bf s+a}) }{(2\pi {\bf i})^MM!}
\sum_{\underline{\lambda}\subseteq M^M } P_{\underline{\lambda}}({\bf s}) P_{\underline{\lambda}^c}({\bf s+a})\,. \notag
\end{align}
Here, ${\rm VdM}({\bf s})$ is the VanderMonde of variables $s_{i=1,\ldots M}$, 
whereas $\underline{\lambda}, \underline{\lambda}^c$ are two conjugate partitions of $M^M$.
From here the residue integration becomes quite more transparent.
First, note that ${\rm VdM}{(\bf s}) P({\bf s)}$ is itself a determinant, 
thus when acted upon by derivatives from ${\tt ibp}$, it vanishes 
in ${\bf s}=0$ unless the derivatives are distributed as  
$\partial_{s_i}^{\#_i}$ with $\#_i=\lambda_i+(i-1)$ or permutations thereof. 
Consider now the $j$th cut. Each  $E_{\mathbf{a}_j}(\mathbf{u}_j)^{L_j+M_j}$ 
energy factor produces $M_j-1+L_j$  derivatives $\partial_{s_i}$ from {\tt ibp} for each $i=1,\ldots, M_j$. 
The minimal set of {\tt ibp} derivatives that the measure can absorb corresponds to
$\underline{\lambda}=\varnothing$, for which $P_{\varnothing}=1$ and  ${\varnothing^c}=[M_j,...,{M_j}]$. 
We find that for each of the $s_{i}$ a number $({M_j-1})/{2}+L_j$ of derivatives
distribute antisymmetrically on $P_{[M_j,...,{M_j}]}({\bf s}_j+{\bf a}_j)\times {\rm VdM} \times \rho_j^{\bf s}
\times {\bf H}_{{\bf a},{\bf b}}\times {\cal T}_{\underline{\bf a}}$. 
From combinatorics it follows that $(\log\rho_j)^{M_j L_j}$
is the maximal power of $\log\rho_j$. The remaining  
 derivatives, in number $({M_j-1})/{2}$ for each of the $s_{i}$, with $i=1,\ldots , M_j$, and for all $j=1,\ldots n-3$, will act on
the rest of the integrand, ie.~factors $P_{[M_j,...,{M_j}]}({\bf s}_j+{\bf a}_j)\times {\rm VdM}$
and ${\bf H}_{{\bf a},{\bf b}}\times {\cal T}_{\underline{\bf a}}$,
and an expression for the maxlog coefficient function follows. A similar 
analysis generalises to $\underline \lambda\neq \varnothing$.

In the special case of the 4pt fishnets 
there is only a cut and so the maxlog is given by ${\tt ibp}$
derivatives acting on $P_{[M_1,...,{M_1}]}({\bf s}+{\bf a})\times{\rm VdM}$.
Hence, a determinantal formula for the maxlog follows immediately by construction; 
and by imposing Steinmann relations as in \cite{Basso:2017jwq} 
one can straightforwardly lift this maxlog determinant and obtain a version of the Basso-Dixon result.\\
\vspace{-1mm}

\begin{figure}[t]
\begin{center}
\includegraphics[scale=0.60]{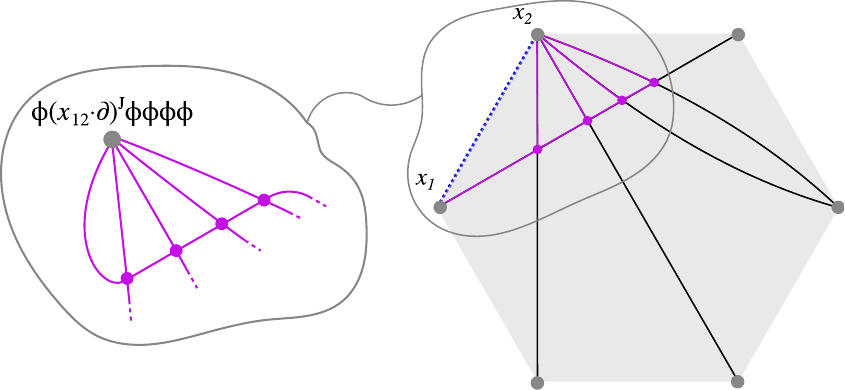}
\caption{The signature $(-,-,-)$ captures explicitly the $\log^4 \rho_1$ behaviour of 
the six-point ladder with $L_{i=1,2}=1$ and $L_3=2$ in the limit $x_{12}^2 \to 0$. 
Here, the displacement between fields at $x_1,x_2$ is replaced by the expansion
$\text{Tr}[\phi(x_1) (x_{12}^{\mu}{\partial}_{\mu})^J \phi(x_1)^4 ]$.}
\label{UV_LL}
\end{center}
\vspace{-8 mm}
\end{figure}

{\bf Stampedes.}  
In planar and light-cone kinematics we can combine SoV 
results with \emph{stampedes} technique. Let $S_j =\{j,\dots j+{m_j}\}$ be a collection 
of consecutive indices, for $j\in P\subset \{1,\ldots,n\}$. We consider
the regime where points in $S_j$ approach a light-ray,
that is $x_{i\in S_j}=(w_{i\in S_j},\bar w_j)$ in light-cone coordinates, 
and look at the  leading behaviour of the diagram. For an $n$-pt fishnet 
we then expect a generalised Taylor expansion of the form
\begin{equation}\label{stampede_ini}
{\cal I}_{M\times L}=\!\!\!\!\!\sum_{\substack{ |\underline k|=M\!L \\[.1cm] \underline{J}\ge 0}}
\left[\left( \prod_{j\in P} \lambda_j^{k_j}\!\!\!\!\prod_{i\in S_j, i\neq j} w_{i,j}^{J_i}\right)
\mathbf{f}_{\underline{k};{\underline J}}({\eta},{\bar \eta}) + \! ...
\right]+...
\end{equation}
where $\lambda_j \sim -\log \bar w_{i,j}$.
The functions $ \mathbf{f}_{k_1,\dots,k_p;\underline J}$ carry the leftover coordinate dependence,
denoted collectively as $(\eta,{\bar \eta})$.
For \eqref{stampede_ini} to hold, 
any vertex of the diagram has to be the intersection 
of two lines emitted from external points $x_i$, $x_k$ 
such that $\exists \,j \in P \,| \,  i,k \in S_j$.

We are going to show that the $ \mathbf{f}_{k_1,\dots,k_p;\underline J}$ are generated via combinatorics. 
First, fields that become null-separated, e.g.~$\phi(x_j), \phi'(x_{j+1})$ for $j \in P$, 
are replaced by a Taylor expansion along a light-ray.
Doing so, we create states $|\text{in} (J)\rangle_{j}$ obtained from inserting light-cone derivatives $\hat \partial^n \equiv \partial^n/n!$ between a number $R$ and $R'$ of fields emitted from the points $x_j$ and $x_{j+1}$, 
\begin{equation}
(\phi_1\cdots \phi_{R})(x_j) \hat \partial^J (\phi'_1\cdots \phi'_{R'})(x_j)|{\rm vacuum}\rangle= |\text{in} (J)\rangle_{j}\\ 
\label{init_exp}
\end{equation}
as in FIG.\ref{UV_LL}.  Last formula becomes $|\text{in} (J_1,J_2,J_3,\dots)\rangle_{j}$ 
when also fields $\phi''(x_{j+2}),\,\phi'''(x_{j+3}),...$ are expanded around $x_j$. Next, we will 
generate loop corrections by acting on these states with the one-loop 
dilations $\mathbb{H}_i$ at position $x_i$. Each $\phi$ is one of the fields 
$ {\cal X}, \bar {\cal X}$ or ${\cal Z}, \bar {\cal Z}$ in the fishnet CFT \cite{Gurdogan:2015csr,Derkachov:2019tzo}, whose propagators 
are rows/columns of the fishnet diagram. The action of $\mathbb{H}_i$ is  
\begin{equation}\notag
\!\mathbb{H}_i\, | \hat \partial^{h_1} \phi_1 \dots\hat \partial^{h_{R}} \phi_{R} \rangle_{i}
=\!\!\!\sum_{k=1}^{R-1} \mathbbm{h}_{k,k+1} \,|\hat \partial^{h_1} \phi_1\dots \hat \partial^{h_{R}} \phi_{R} \rangle_{i}
 \vspace{-3 mm}
\end{equation}
where
 \begin{equation}
 \vspace{-1 mm}
\mathbbm h_{12} | \hat \partial^{h_1} {\cal X},\hat \partial^{h_2} {\cal Z} \rangle_i 
= \! \sum_{j+l=h_1+h_2}^{} \frac{|\hat \partial^{j} {\cal Z},\hat \partial^{l} {\cal X}\rangle_i}{h_1+h_2+1}\,.
\end{equation}
Other non-zero matrix elements are obtained by cyclic permutations 
${\cal X}\!\to\! {\cal Z} \!\to \! \bar{\cal X}\!\to\!\bar{\cal Z}$. In particular, states 
with opposite \emph{chirality} are annihilated
$\hat h_{12}\, | \hat\partial^{n} {\cal Z} , \hat\partial^{m} {\cal X} \rangle =0$.

At this point, $\mathbf{f}_{k_1,\dots,k_p;J}$  is the monomial of degree $(k_1,\dots,k_p)$ 
in the ``times" $\lambda_j$, of the generating function
\begin{align}
\label{generating_function}
&
\mathcal{G}_{\underline J}(\lambda_j) = \mathbf{C} \cdot e^{\sum_{i \in P} \lambda_i \mathbb{H}_i} 
|\text{in}\left(\underline J\right)\rangle\,, \\[.2cm]
&
|\text{in}\left(\underline J\right)\rangle=
\bigotimes_{ \substack{ i \in \{1,\dots,n\}\backslash S\\[.15cm] S\equiv \cup_j S_j \backslash \{{j}\} }} 
\!\!\!\!|\text{in}\left(\underline{J_i}\right) \rangle_{i}\,  \notag
\end{align}
The initial state is given by \eqref{init_exp} 
whenever  $i \in P$,
\begin{equation*}
|\text{in}\left(\underline{J_i}\right) \rangle_{i} \equiv  |\text{in}(J_{h_1},\dots ,  J_{h_{m_i-1}}) \rangle_{i}\,, \vspace{-1 mm}
\end{equation*}
otherwise $\underline{J_i}=\underline 0$ for $i\notin P$.
Finally, the functional $\mathbf{C}$ is the free-theory contraction of fields 
$\phi_i(y_i)$ and $\phi_i'(y'_i)$
 that stand at endpoints of a given line in the fishnet lattice.
\begin{figure}[t]\vspace{-1.5 mm}
\includegraphics[scale=1.1]{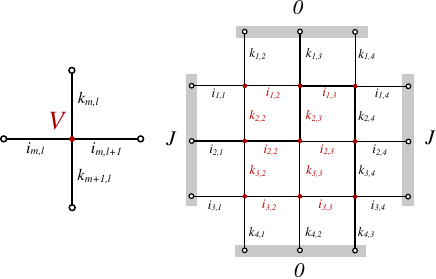}\vspace{-1.5 mm}
\caption{The vertex model for $\mathbf{f}_{ML,J}$ with $M=L=3$. Indices 
$i_{m,l}$ and $k_{m,l}$ are summed over in the bulk, and constrained along 
boundaries. Vertically, $\sum_{m=1}^M i_{m,1} = J=\sum_{m=1}^M i_{m,L+1}$. 
Horizontally $k_{1,l}=k_{M+1,l}=0$.}
\label{pf_BD}
\end{figure}

For example, take a $4$-pt fishnet with $S_1\!=\!\{1,2\}$; 
 \begin{equation}
 \label{stampede_4pt}
\mathbf{f}_{ML, J} = \frac{1}{\prod_{h=1}^L\prod_{k=1}^M (M+L-h-k+1)} \times \mathbb{F}_{M,L}(J)\,,
\end{equation} 
where $\mathbb{F}_{M,L}(J)$ is the canonical partition function of a vertex model on 
a square-lattice of size $M\times L$,  with $J$ excitations, Boltzmann weights  \vspace{-2.5 mm}
 \begin{equation*}
V(a,b;c,d)= \frac{\delta_{a+b,c+d}}{a+b+1}\,, \vspace{-2 mm}
\end{equation*} 
and domain-wall boundary conditions as in FIG.\ref{pf_BD}.
From this description, the maxlog in \eqref{stampede_ini} 
is the gran-canonical partition function of the above vertex-model 
with fugacity $Z=w_{12}w_{34}/w_{14}w_{23}$.
The \emph{zero-temperature} pre-factor in \eqref{stampede_4pt} 
was computed in \cite{Olivucci:2021pss} as the euclidean OPE limit $Z \to 0$.
The same vertex model describes maxlogs for higher-pt partition functions.\\
\vspace{-1mm}

{\bf Conclusions and Outlook.}
We initiated a systematic study
of multi-point Feynman integrals at any-loop order in planar kinematics,
by combining integrability/SoV methods \cite{Basso:2018cvy, Derkachov:2019tzo, Olivucci:2021cfy, Olivucci:2023tnw}, 
with Mellin space and symmetric polynomials techniques. 
We first solved a class of point-split ladder integrals by an elegant recursion, 
and then for $n$-point fishnets, we introduced a Cauchy identity representation 
of the SoV measure, showing how this could be used 
to understand the pattern of emergent structures.  

Some future directions. Our results apply directly to Fishnet CFTs, 
a family of integrable 4d theories closely related to ${\cal N}=4$ SYM
\cite{Gurdogan:2015csr, Caetano:2016ydc, Gromov:2017cja},
where it is known that
all Feynman diagrams 
ultimately should be related to partition functions of an integrable lattice model  \cite{Zamolodchikov:1980mb}.  
Throughout our discussion we have indeed provided concrete realisations of this statement. 
But in addition, fishnet diagrams describe conformal correlators and can be considered as a simplified playground 
to familiarise with CFT/integrability tools for multi-point correlators, eg.~\cite{Rosenhaus:2018zqn,Buric:2020dyz,Poland:2021xjs,Buric:2021ywo,Buric:2021ttm,Buric:2021kgy,Kaviraj:2022wbw}.
In this respect, beyond one-loop,
not much is known about multi-point conformal integrals and the space of function they describe,
see \cite{Chicherin:2015edu,Chicherin:2017frs, Hannesdottir:2022xki,McLeod:2023qdf, McLeod:2023doa} for recent progress.
An intriguing direction would be to bootstrap the diagrams we discussed here
by combining 1) data from the SoV log stratification, 
2) an ansatz in terms of pure functions (such as 
\cite{Goncharov:2001iea,Schnetz:2013hqa, Panzer:2015ida, Broedel:2019hyg,tesi_James}) and 3) analytic constraints (single-valuedness, Steinmann relations \cite{Steinmann,Caron-Huot:2016owq}, etc.).
We expect integrability to imply simple structures at any loop order, generalising 
non trivially the determinants found by \cite{Basso:2018cvy}. 
Finally, it would be interesting to  lift our methods out of the plane, and  
extend the use of SoV technique to multi-point conformal integrals 
with other topologies, but similar analytic properties.\\

{\bf Acknowledgments.}
We thank James Drummond for discussions and for providing us the symbol of ${\cal L}_{1,1}$, 
and \"Omer G\"urdogan for helping us integrating the symbol. 
We also thank Frank Coronado, Paul Heslop and Pedro Vieira for very helpful comments.  
Research at Perimeter Institute is supported in part by the Government of Canada 
through the Department of Innovation, Science, and Economic
Development Canada and by the Province of Ontario through the Ministry of Colleges
and Universities. 
FA is supported by the Ramon y Cajal program through the fellowship RYC2021-031627-I funded 
by MCIN/AEI/10.13039/501100011033 and by the European Union NextGenerationEU/PRTR. 
Finally, we acknowledgment the FAPESP grants 2019/24277-8 and 2020/16337 for support during our stay at the ICTP
South American Institute for Fundamental Research, IFT-UNESP, in Sao Paulo,  
where part of this work was done.

\newpage

\onecolumngrid  \vspace{1cm} 
\begin{center}  
{\Large\bf Supplementary Material} 
\end{center} 
\tableofcontents

\section{Our conventions}

In this section we describe our conventions on the measure $\mu$, the interaction term $H_{a,b}({u},{ v})$, and the $R$-matrix.\\

{\bf Measure.} This is given by
\begin{equation}\label{defmu}
\mu_{{\bf a}}({\bf u})= \frac{1}{(2\pi)^M 
 M!} \prod_{k=1}^{M} a_k \prod_{h=k+1}^M \left( (u_h-u_k)^2+\frac{(a_h-a_k)^2}{4}\right)\left( (u_h-u_k)^2+\frac{(a_h+a_k)^2}{4}\right)
\end{equation}
\\

{\bf Interaction term.} This is given by
\begin{equation}\label{defHab}
H_{ab}(u,v)= (-)^{b-1}\,  \frac{ \Gamma[1+\frac{a}{2}+{\bf i} u]\Gamma[1+\frac{b}{2}-{\bf i}v]}{\Gamma[\frac{a}{2}-{\bf i}u]\Gamma[\frac{b}{2}+{\bf i}v]}
\frac{\Gamma[\frac{a-b}{2}-{\bf i}u+{\bf i} v]}{\ \Gamma[1+\frac{a+b}{2}+{\bf i}u -{\bf i}v ]\ } =\widetilde H_{ab}(u,v)\times \frac{\Gamma[1+\frac{a-b}{2}+{\bf i}u -{\bf i}v]}{ \Gamma[\frac{a+b}{2}+{\bf i}u-{\bf i}v]}
\end{equation}
The function $\widetilde{H}$ coincides with the definition given in \cite{Olivucci:2023tnw}, 
up to the following re-labelling of bound-state index and rapidities $(a,b) \to (n+1,m+1)$ and $(u,v) \to (\nu+i/2,\mu+i/2)$. 
We changed normalization from $\tilde H$ to $H$ since we found it more convenient 
when integrating our SoV integrand as a Mellin space integral. ~\\

{\bf R-matrix.} Given two representations $V_{i=a,b}$ of $su(2)$, with ${\rm dim}(V_{j})=j$, 
then $R_{a,b}$ belongs to ${\rm End}(V_a\otimes V_b)$.
When one of the two representations is the fundamental and the other 
is arbitrary, the $R$-matrix is known explicitly. 
In our conventions, see \cite{Maillet:2019nsy}, we have:
\begin{align}
{R}_{a,b=1 }(u)=&\ \ 1_a\qquad \forall a\ge 1\\[.2cm]
{R}_{a,b=2 }(u)=&  
\left[ \begin{array}{cc}   (u+\frac{1}{2})1_a  + J_{3;a} & J_{-;a}   \\[.2cm]  J_{+;a} & (u +\frac{1}{2})1_a -J_{3;a} \end{array} \right] \qquad \forall a\ge 1
\label{Rab_ini}
\end{align}
with $1_a$ the identity matrix, and $J_{\pm,3;a}$ generators of $su(2)$:  
\begin{align}
\big[ J_3\big]_{1\leq i,j\leq a}=\delta_{i,j} \times\, \tfrac{(a-2i+1)}{2}\qquad;\qquad
\big[ J_{\pm}\big]_{1\leq i,j\leq a}=&\ \delta_{i,j\mp 1} \times \sqrt{ (j-\tfrac{1\pm 1}{2})(a-j+\tfrac{1\pm1}{2}) } 
\end{align}
Note, when $a=1$ there is only the identity matrix. \\
%

To grow the representation $b>2$ we will use the fusion formalism of \cite{Kulish:1981gi}. 
Fusion implies that
\begin{equation}
{R}_{a,b\ge2}(u)= \frac{1}{(b-1)!} \sum_{\sigma} [{R}_{a,2}(u-\tfrac{b-2}{2})]_{l_1 r_{\sigma(1)} } . [ {R}_{a,2}(u-\tfrac{b-2}{2}+1)]_{l_2 r_{\sigma(2)}} \ldots [{R}_{a,2}(u+\tfrac{b-2}{2})]_{l_{b-1} r_{\sigma(b-1)}} \,.
\end{equation}
%
%
Here each entry $[R_{a,2}]_{1\leq l,r \leq 2}\in {\rm End}(V_a)$ and the ``dot'' product 
stands for ordinary matrix multiplication. 
Fusion returns a $2^{b-1}\times 2^{b-1}$ matrix where the ``\,$2$\," is the fundamental of 
$su(2)$ we started with. But the rank of $R_{ab}$ as a matrix on the space $V_b$ 
is only $b$. To make this manifest we can change basis: 
out of the $2^{b-1}$ basis elements parametrised as $|e_1,\ldots e_{b-1}\rangle$ 
with $e_j=\{1,0\}$ for each $j=1,\ldots,b-1$,
we pick the $b$ elements obtained by symmetrising the combinations
$|11\ldots 11\rangle,|11\ldots 10\rangle,\ldots ,|10\ldots 00\rangle,|00\ldots 00\rangle$. 
Then we consider a Gram-Schmidt orthogonalisation, with these $b$ elements fixed, of the $|e_1,\ldots e_{b-1}\rangle$.
In the new basis $R_{ab}$ is non trivial only in a $b\times b$ block. \\

Comparing with the literature \cite{Maillet:2019nsy}, 
\begin{equation}\label{normaliz_b2}
\widetilde{R}_{a,2 }(u)=\frac{ {R}_{a,2 }(u) }{u+\frac{1}{2}+\frac{a-1}{2} }\qquad;\qquad
\widetilde{R}_{a,b}(u)=\frac{{R}_{ab}(u)}{(u+\frac{a-b}{2}+1)_{b-1} }\,.
\end{equation}
In our conventions, ${R}_{a,b}(u)$ is the polynomial part, in $u$, of ${\tilde R}_{a,b}(u)$. The normalization of  ${\tilde R}_{a,b}(u)$ 
is fixed so that $\tilde{R}_{a,b}(u)\tilde{R}_{a,b}(-u)=1$.
Equivalent formulae 
can be obtained starting from $R_{a=2,b}$ and fusing on the first space. \\ 


In the main text we introduced $R_{a_k b_h}$ in ${\rm End}(\otimes_{j,m} V_{{\tt s}_{jm}})$.
This generalisation is quite straightforward. Say we consider $V_{a_1}\otimes\ldots\otimes V_{a_k}\otimes \ldots \otimes V_{b_1}$,
the matrix $R_{a_k b_1=2}$ is the $2\times 2$ matrix defined as in \eqref{Rab_ini}, 
where the entries now are tensored with the identity in all other spaces that are not $V_{a_k}$. 
By fusion we can grow the representation $b_1\ge 2$. If we add more spaces on the right of 
$V_{b_1}$, again we tensor with the identity on the right. 
As another example, define 
${\bf R}_{a_1\ldots a_M,b_1}\equiv\prod_{k=1}^M R_{a_kb_1}$. Then, ${\bf R}_{a_1\ldots a_M,b_1=2}$ is 
obtained by matrix multiplication of $2\times 2$ matrices $R_{a_k,b_1=2}$ described above. 
In particular, the entries belong to ${\rm End}(\otimes_{k=1}^M V_{a_k})$.
By fusion we can grow the representation $b_1\ge 2$.

\section{SoV representation of Fishnets integrals on the plane}


In this section we present a concise derivation of the SoV representation in \eqref{calFdef}. 
Our formula is obtained from a more general SoV representation {\cite{Derkachov:2019tzo,Olivucci:2021cfy,Olivucci:2023tnw} }
when we restrict the kinematics to the plane. For a point $x^{\mu}$ this restriction 
is conveniently parametrized by a pair of light-cone coordinates $(w,\bar w)$,
\begin{equation}
\label{4d_to_2d}
x^{\mu} = (x_0,x_1,x_2,x_3) \,\xrightarrow{\text{on the plane}} \, \left( \frac{w+\bar w}{2},0,0,\frac{w-\bar w}{2 i }\right)\,.
\end{equation}
In euclidean signature the pair $(w,\bar w)$ identifies a complex number of 
modulo $|x|^2 = w \bar w$ and phase ${w}/{\bar{w}}=e^{2{\bf i} \theta}$. We associate to 
this angle a corresponding rotation on the plane in the representation $V_a$ of $SU(2)$, as follows 
\begin{equation}
\left(\frac{w}{\bar w} \right)^{J_{3;a}} =e^{2 {\bf i} \theta J_{3;a}} = \left(\cos \theta\, 1_2 + {\bf i} \sin \theta \sigma_3\right)^{\otimes (a-1)}\,.
\end{equation} 
The latter is nothing but the reduction of 
\begin{equation}
\label{x_su2}
[\mathbf{x}]^{a-1} \equiv \left(\frac{\ x^{\mu}}{|x|}\, \mathbf i{\sigma}_{\mu}\right)^{\otimes (a-1)} \,\xrightarrow{\text{on the plane}} \, \left(\frac{w}{\bar w} \right)^{J_{3;a}}\,.
\end{equation}
where the matrix $\mathbf{x}$ belongs to $SU(2)$ and it is related to the coordinates 
$x^{\mu}$ in \eqref{x_su2} via the Pauli matrices $\sigma_{k=1,2,3}$ and 
$\sigma_0=I_{2}$. This type of matrices enter the SoV formalism as a building-block  
which contains the rotational part of the fishnet `mirror' eigenfunctions \cite{Derkachov:2019tzo}. 
The reduction to planar kinematic simplifies considerably the SoV representation 
as it replaces the general $SU(2)$ matrix $[\mathbf{x}]$ with a $U(1)$ rotation.
%
%
Let us show how this works in more detail.\\
 \begin{figure}[t]
\includegraphics[scale=0.65]{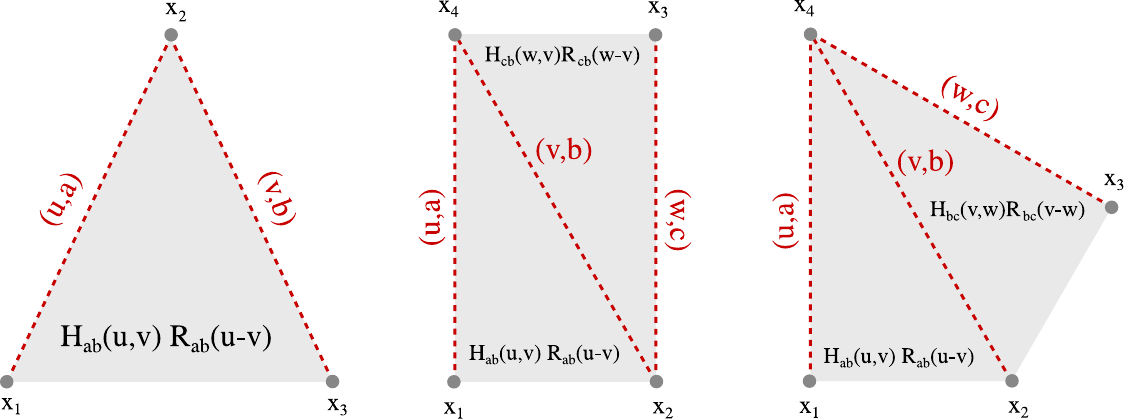}
\caption{\textbf{Left:} A tile of the SoV triangulation of a fishnet integral carrying 
excitations $(\mathbf{u},\mathbf a)$ and $(\mathbf{v},\mathbf b)$ on its cuts 
(red dashed edges). \textbf{Middle:} The quadrilateral obtained by gluing two 
triangles flipped with respect to each other. \textbf{Right:} The quadrilateral obtained by gluing two 
triangles with the same \emph{tip} (ie the vertex common to two cuts).  }
\label{glued_square}
\vspace{-4 mm}
\end{figure}

\begin{figure}[b]
\begin{center}
\includegraphics[scale=0.48]{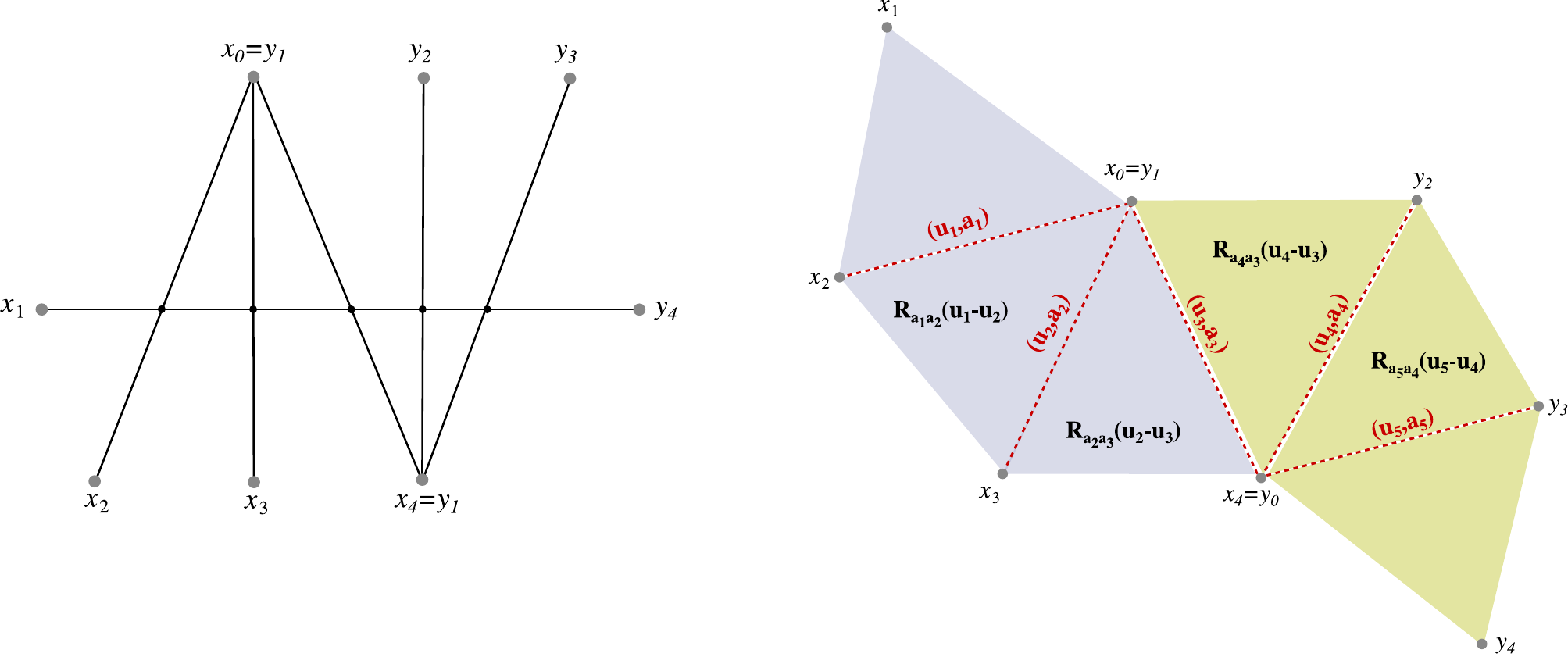}
\caption{An example of 8pt split-ladder diagram where the point-splitting takes place at different points $x_0,y_0$ (left) and its SoV triangulation (right). We label the cuts by the excitations $(u_k,a_k)$ inserted each of them. We distinguish a blue and a yellow region made both of $m=m'=3$ triangular tiles glued sequentially, with the tip in common. The two regions are then glued to each other along one edge $(u_3,a_3)$.}
\label{snake}
\end{center}
\end{figure}
Each triangular ``tile" of a fishnet integral, see FIG.\ref{glued_square} (left), contributes to the SoV integrand with the following matrix-part, 
\begin{equation}
 [\mathbf{x}_{13}\overline{\mathbf{x}_{32}}]^{\mathbf a-1}
\mathbf R_{\mathbf{ab}}(\mathbf{u}-\mathbf{v})  [\mathbf{x}_{21}\overline{\mathbf{x}_{13}}]^{\mathbf b-1}\,.
\end{equation}
Assembling adjacent tiles of the triangulation means that we \emph{glue} them along common \emph{cuts}. 
At first, one forms a quadrilateral out of two triangles, as exemplified in FIG.\ref{glued_square} (middle, right), 
with two possible relative orientations. Then, step-by-step one reconstructs the full $n$-gon that contains 
the $n$-point fishnet integral. 
The general scheme for the gluing procedure is the following. First, the $n$-gon 
can be divided into a number of subregions, each one made of a sequence of say $m$ triangles glued along $m-1$ 
edges in the way of FIG.\ref{glued_square} (right), i.e. having the \emph{tip} in common. For example, in the triangulation of the 8pt 
split-ladder of FIG.\ref{snake} these two subregions are coloured in blue and yellow. Then, different subregions 
are glued with each other along an edge, in the way of FIG.\ref{glued_square} (middle).\\ 

Now, we can write the expression for a general SoV integrand, i.e.~when more subregions are present. The derivation is based on the rules of \cite{Olivucci:2023tnw}, further
specialized to the planar kinematics \eqref{4d_to_2d}. In order to understand the general case, it will be enough to illustrate the gluing procedure in presence of two subregions.
Consider first a subregion of $m$ triangles 
having in common the tip $x_0$, eg. the blue region in FIG.\ref{snake}. This contributes to 
the matrix-part of the integrand via the following product of matrices of type $[\mathbf x ]$ and $[\bar{\mathbf{x}}]\equiv[\mathbf{x}]^{\dagger}$, and $R$-matrices
\begin{equation}
\label{region}
\otimes_{k=1}^{m-1}  [\overline{\mathbf{x}}_{0,k+2} \mathbf{x}_{k+2,k+1}\overline{\mathbf{x}}_{k+1,k} \mathbf{x}_{k 0}]^{\mathbf a_k-1} \times \prod_{k=2}^{m-1}\mathbf R_{\mathbf{a_{k-1} a_{k}}}(\mathbf{u}_{k-1}-\mathbf{u}_k) \,.
\end{equation}
The $n$-pt split fishnet integrals analyzed in this letter are defined 
via a sequential triangulation, see FIG.\ref{slicing}, featuring a number $m=n-2$ of triangles with common tip. 
%
%
In this case the SoV representation takes the form \eqref{T_parition}, 
and follows from the reduction of \eqref{region} to the plane, ie. \eqref{x_su2}, and finally expressing 
the cross ratios defined for every quadrilateral according to defs \eqref{crossratios}. But in the general 
case, the triangulation contains more subregions, and the SoV integrand has a more varied shape. 
 For example,  in FIG.\ref{snake} there are two subregions of
size $m=m'=3$, glued along one cut, $x_{04}$. 
The blue region contributes with \eqref{region}, while the yellow region yelds
\begin{equation}
\label{regions_glued_II}
  \otimes_{k=1}^{m'-1}  [\overline{\mathbf{y}}_{0k}{\mathbf{y}_{k+1}}\overline{\mathbf{y}}_{k+1k+2}{\mathbf{y}_{k+20}}]^{\mathbf a_{k+m}-1}  \times \prod_{k=m+m'-1}^{m+2}\mathbf R_{\mathbf{a_{k} a_{k-1}}}(\mathbf{u}_{k}-\mathbf{u}_{k-1})\,.
\end{equation}
Then, these two contributions multiply from left/right respectively the term involving $(\mathbf u_m, \mathbf a_m)$, ie. the excitations inserted along the cut that connects the two regions
\begin{equation}
\label{connection}
 \mathbf R_{\mathbf{a}_{m-1} \mathbf{a}_{m}}(\mathbf{u}_{m-1}-\mathbf{u}_{m})  [\overline{\mathbf{y}}_{12}{\mathbf{y}_{20}}]^{\mathbf a_{m}-1} \mathbf R_{\mathbf{a}_{m+1} \mathbf a_{m}}(\mathbf{u}_{m+1}-\mathbf{u}_{m}) [\overline{\mathbf{x}}_{m+1m}{\mathbf{x}_{m0}}]^{\mathbf a_{m}-1}\,.
\end{equation}
The contribution to the integrand is the trace over all representations $V_{\mathbf a_k}$ 
of the product \eqref{region}$\cdot$\eqref{regions_glued_II}$\cdot$\eqref{connection}. 
It shall be clear now that this scheme iterates in the general case of a fishnet integral 
composed by two or more subregions.
The reduction of the matrix part to $2d$ kinematics is easy to express 
via conformal invariants, following \eqref{x_su2} with the cross ratios defined in \eqref{crossratios}; 
indeed, it amounts to the identifications
\begin{equation}
[\overline{\mathbf{x}}_{0k+2} \mathbf{x}_{k+2k+1}\overline{\mathbf{x}}_{k+1k} \mathbf{x}_{k 0}]^{a-1} = \left(\frac{Z_{k}}{\bar Z_{k}}\right)^{J_{3; a}}\;\;\;\;\;\,;\;\;\;\;\;\;\;\;  [\overline{\mathbf{y}}_{0k}{\mathbf{y}_{k+1}}\overline{\mathbf{y}}_{k+1k+2}{\mathbf{y}_{k+20}}]^{a-1}=\left(\frac{Z_{k+m}}{\bar Z_{k+m}}\right)^{J_{3; a}}\,.
\end{equation}
 On the other hand, the connecting term in the form \eqref{connection} is not a manifest 
 function of conformal invariants. In order to reach the final, $Z,\bar Z$-dependent form of 
 the SoV integrand, one shall rearrange the terms in the matrix part. First, we use the 
 unitarity of matrices $[\mathbf{x}]^{a-1}$, and the $SU(2)$-invariance of the $R$-matrix, expressed as
 $R_{a,b}(u)[\mathbf{x}]^{a-1}[\mathbf{x}]^{b-1} = [\mathbf{x}]^{a-1}[\mathbf{x}]^{b-1} R_{a,b}(u)$, on equation \eqref{connection}. By doing so, the ratio $Z_m/\bar Z_m$ can be made manifest 
 and \eqref{connection} can be recasted as follows
 %
 %
\begin{equation}
 \mathbf R_{\mathbf{a}_{m-1} \mathbf{a}_{m}}(\mathbf{u}_{m-1}-\mathbf{u}_{m}) [\overline{\mathbf{y}}_{20}{\mathbf{y}_{12}}]^{\mathbf a_{m+1}-1}   \mathbf R_{\mathbf{a}_{m+1} \mathbf{a}_m}(\mathbf{u}_{m+1}-\mathbf{u}_{m}) [\overline{\mathbf{y}}_{12}{\mathbf{y}_{20}}]^{\mathbf a_{m+1}-1}   \left(\frac{Z_{m}}{\bar Z_{m}}\right)^{J_{z;\mathbf a_{m}}} \,.
\end{equation}
In order to get rid of the spurious terms of type $[\mathbf x]^{\mathbf a_{m+1}-1}$ in this expression, 
we insert the identity as a product of spurious factors in each of the representation spaces 
that stand on the right of $V_{\mathbf a_{m+1}}$ in the matrix product, that is
\begin{equation}
(32)\cdot (33) =  (32)  \cdot \left(\otimes_{k=m+2}^{m'-1} [\overline{\mathbf{y}}_{20}{\mathbf{y}_{12}}]^{\mathbf a_{k}-1} [\overline{\mathbf{y}}_{12}{\mathbf{y}_{20}}]^{\mathbf a_{k}-1} \right)  \cdot (33)\,.
\end{equation}
Next, we make use of planar kinematics, since the rotations  $[\mathbf{x}]^{a-1}$ 
on the plane become abelian, and in particular commute with matrices $(Z/\bar Z)^{J_{3;a}}$. 
Hence, we push the spurious factors to the sides of \eqref{regions_glued_II}$\cdot$\eqref{connection}, to get to
\begin{equation}
\left(\otimes_{k=m+1}^{m'-1} [\overline{\mathbf{y}}_{20}{\mathbf{y}_{12}}]^{\mathbf a_{k}-1}\right)  \left(\otimes_{k=m}^{m'-1}\left(\frac{Z_{k}}{\bar Z_{k}}\right)^{J_{3;\mathbf a_{k}}} \times\prod_{k=m+m'-1}^{m+1} \mathbf R_{\mathbf a_{k} \mathbf a_{k-1}}(\mathbf{u}_k-\mathbf{u}_{k-1})\right)  \left( \otimes_{k=m+1}^{m'-1} [\overline{\mathbf{y}}_{12}{\mathbf{y}_{20}}]^{\mathbf a_{k}-1} \right)\,.
\end{equation}
Under the trace $\text{Tr}_{\mathbf{a}}$ the spurious factors cancel by unitarity and the final contribution to the SoV integrand reads
\begin{equation}
 \text{Tr}_{\mathbf{a}_1}\!\!\cdots  \text{Tr}_{\mathbf{a}_{n-3}} \left[\otimes_{k=1}^{m+m'-1}  \left(\frac{Z_{k}}{\bar Z_{k}}\right)^{J_{3;\mathbf a_{k}}} \times \prod_{k=1}^{m-1}  \mathbf R_{\mathbf a_{k} \mathbf a_{k+1}}(\mathbf{u}_k-\mathbf{u}_{k+1}) \times \prod_{k=m+m'-1}^{m+1}  \mathbf R_{\mathbf a_{k} \mathbf a_{k-1}}(\mathbf{u}_k-\mathbf{u}_{k-1}) \right] \,.
\end{equation}
Last formula distinguishes the two subregions of the diagram via a $\pm$ sign in 
the difference of rapidities that enter, and in the order of $R$-matrices in the product.
With reference to FIG.\ref{snake}, the triangulation of a split-ladder is made of (blue) 
tiles with the tips pointing up -- $x_0$ in the figure -- and tiles (yellow) with 
the tips pointing down -- $x_0'$ in the figure. A generic point-splitting 
is encoded by an $t$-uple of integers $\mathbf{m}= (m_1,\dots,m_{t})$ where $t$ is the 
number of subregions of the diagram, and $m_j$ the number of cuts in that subregion. 
Starting with a region that points up, an odd value of $j$ labels a regions with the tips of triangles that \emph{point up} (eg. blue region), and even one 
labels to a region with the tips that \emph{point down} (eg. yellow region).\\

The SoV integrand for a general Fishnet on a planar $n$gon is given by the following expression 
 \begin{align}
\label{SoV_representation_general}
&
\mathcal{F}_{L_1\ldots L_{n-3},M_1,\dots,M_{n-3}}(\mathbf{m})= \\
&
\ \prod_{j=1}^{n-3}{ \mu}_{ {\bf a}_j}({\bf u}_j) \times E_{\mathbf{a}_j}(\mathbf{u}_j)^{L_j+M_j} \times  
\left(\rho_j\right)^{{+}\sum_{k=1}^{M_j} \left({\bf i} u_{jk} -\frac{1}{2}\right)} 
\times  \prod_{k=1}^{t} \prod_{j=1}^{m_k}{\bf H}_{{b}^{(k)}_{j-1},{b}^{(k)}_{j}}({\bf v}^{(k)}_{j-1}-{\bf v}^{(k)}_{j}) \times 
 {\cal T}_{ \underline{\bf a}}\left[\mathbf{m};\theta_1,\ldots; {\bf u}_1,\ldots \right]  \notag
\end{align}
where $\mathbf{b}^{(k)}$ are the vectors of bound-state indices and $\mathbf{v}^{(k)}$ are rapidities that sit on the edges of the $k$-th region. They are ordered from left to right for odd $k$ and flipped for even $k$, say $(\mathbf{b}^{(1)},\mathbf{v}^{(1)}) = \{(\mathbf a_1,\mathbf u_1),(\mathbf a_2,\mathbf u_2),(\mathbf a_3, \mathbf u_3)\}$ and $(\mathbf{b}^{(2)},\mathbf{v}^{(2)})=\{(\mathbf a_5, \mathbf u_5),(\mathbf a_4,\mathbf u_4),(\mathbf a_3,\mathbf u_3)\}$ in FIG.\ref{snake}. The matrix part of last formula is
\begin{align}\label{T_parition}
{\cal T}_{\underline{\mathbf{a}}}[\mathbf{m}]=\  \text{Tr}_{\mathbf{a}_1}\!\!\cdots  \text{Tr}_{\mathbf{a}_{n-3}}
 \left[\otimes_{j=1}^{n-3} \left(\frac{Z_j}{\bar{Z}_j}\right)^{\!\!{J}_{3; \mathbf a_{j} }}\!\cdot \prod_{k=1}^{t} \ {\bf R}^{(m_k)}\right]
\qquad;\qquad {\bf R}^{(m_k)}= \prod_{l=2}^{m_k}  {\bf R}_{{b}^{(k)}_{l-1},{b}^{(k)}_{l}}({v}^{(k)}_{l-1}-{v}^{(k)}_{l})
\,.\notag
\end{align}

\section{Recursion relations for the $n$-pt split-ladders in $(-,\ldots,-)$ signature}

In this section we explain how to perform the integration of ${\cal L}_{L_1,\ldots, L_{n-3}}$ algorithmically.
First, let us make explicit the SoV integrand for example at 5-pts
 \begin{equation}
 \label{exa_text_5pt}
{\cal L}_{L_1L_2}=
\frac{\rho_1^{s_1}\rho_2^{s_2} }{ \Gamma[1-s_1]  }
\frac{\Gamma[s_2-s_1]\Gamma[-s_2]}{\ \ \ (-s_1)^{L_1} \, (-s_2)^{L_2}}\times \frac{ {\cal T}_{a_1a_2}}{\prod_i(a_i+s_i)^{L_i}}\,,
\end{equation}
and 6-pts,
\begin{equation}
\label{exa_text_6pt}
{\cal L}_{L_1L_2L_3}=
 \frac{\rho_1^{s_1}\rho_2^{s_2} \rho_3^{s_3}}{ \Gamma[1-s_1]  } 
\frac{  \Gamma[s_2-s_1]   \Gamma[s_3-s_2]\Gamma[-s_3]  }{
(-s_1)^{L_1} \,(-s_2)^{L_2}  (-s_3)^{L_3} }\times \frac{ {\cal T}_{a_1a_2a_3} }{ \prod_i(a_i+s_i)^{L_i}}\,.
\end{equation}
It is clear that the integrand sequentiates with the number of external points. Thus,
our strategy is to integrate iteratively starting from $s_{n-3}$ down to $s_2$, and finally $s_1$.
Since $s_j\in\{0,p_j=\mathbb{N}\}$ there are $2^{n-3}$ types of residues that we can take, 
which can be labeled by sequences -- to be read from right to left --
eg. $(s_{1}=0,s_{2}=0,\ldots,s_{n-3}=0)$, $(s_{1}=\mathbb{N},s_{2}=0,\ldots,s_{n-3}=0)$, etc \ldots. 
For example, at 5pt there are four sequences, at 6pt there are eight.\\

In order to exhibit the recursion clearly, we will use $m+1,m,m-1$ 
as labels. We reason as follows. Suppose we performed a number of integrations, 
and we look at the integrand $I_m$ in the variable $s_{m\ge 2}$. 
The result is made of a sum of integrands, all of the form
\begin{equation}
\label{ini_rec}
I_{i_mj_m}^{\,q_{m+1}}
=\prod_{k=2}^{m-1}\frac{ \Gamma[s_k-s_{k-1}]}{(-s_k)^{L_k}}\ \partial_{s_{m}}^{ {\color{blue} j_m} }\Gamma[s_{m}-s_{m-1}]
\frac{ \Gamma[q_{m+1}-s_{m}] }{ \ \ (-s_{m})^{L_{m}+{\color{blue} i_m}}  }\, {\rm RHS}^{\,q_{m+1}}_{i_mj_m}\qquad;\qquad m\ge 2\,.
\end{equation}
There are two pieces of \emph{data} to be considered here: the possible values 
of the integers $i_m,j_m$ in {\color{blue} blue}, and the value of $q_{m+1}\in\{0,p_{m+1}=\mathbb{N}\}$ 
which comes from the integration of $s_{m+1}$.
The RHS we take it as given, and label it with the data we just mentioned.
At this point, we perform the integration in $s_m$, 
therefore we need to decide which poles in $s_n$ we want to focus on. 
We can pick $s_m\in\{0,p_m=\mathbb{N}\}$. Depending on this choice, 
we generate a set of ${\rm RHS}_{i_{m-1}j_{m-1}}$ for various values of $i_{m-1},j_{m-1}$.  
It turns out that when take the residue in $s_m=0$ we find
\begin{equation}\label{inzerorec}
 \underbracket{ {\rm  RHS}_{i_{m-1}j_{m-1}}^{\ 0 \rightarrow 0} }_{\rule{0cm}{.4cm} 0\,\leq\, i_{m-1}+j_{m-1}\leq L_m+i_m+j_m}\qquad;\qquad
 \underbracket{ {\rm  RHS}_{i_{m-1}j_{m-1}}^{\ p_{m+1} \rightarrow 0} }_{\rule{0cm}{.4cm} 0\,\leq\, i_{m-1}+j_{m-1}\leq L_m-1+i_m+j_m}
 \end{equation}
and when we take the residue in $s_m=p_m$ we find,
 \begin{equation}\label{inprec}
  \underbracket{ {\rm  RHS}_{i_{m-1}j_{m-1}}^{\ 0 \rightarrow p_m} }_{\rule{0cm}{.4cm} 0\,\leq\, i_{m-1}+j_{m-1}\leq j_m}\qquad;\qquad
  \underbracket{  {\rm  RHS}_{i_{m-1}j_{m-1}}^{\ p_{m+1} \rightarrow p_m} }_{\rule{0cm}{.4cm} 0\,\leq\, i_{m-1}+j_{m-1}\leq j_m}
\end{equation}
For orientation, consider ${\cal L}_{L_1L_2}$. The starting point of the recursion is 
the identification of $I^{q_3=0}_{i_2=j_2=0}$, which we read directly from
SoV representation, therefore from \eqref{exa_text_5pt}. This is 
\begin{equation}
\textnormal{5-pt}\qquad;\qquad {\rm RHS}_{i_2=j_2=0}^{\,q_3=0}=
\widetilde{\cal T}_{a_1a_2}\equiv \frac{  \rho_1^{s_1}\rho_2^{s_2} {\cal T}_{a_1a_2}(s_1,s_2) }{ (a_1+s_1)^{L_1}(a_2+s_2)^{L_2} }\,,
\end{equation}
Similarly, for any number of points, the starting point is read off directly from the SoV representation.\\


The recursive step \eqref{inzerorec} -- generated by considering the residue in $s_m=0$ -- gives 
\begin{mdframed}
\begin{align}\label{rec_for0}
&{\rm RHS}_{i_{m-1}j_{m-1}}^{\ q_{m+1}\rightarrow 0}\Big[i_m,j_m,{\rm RHS}_{i_mj_m}\Big]=  
\frac{ (L_{m-1})_{i_{m-1}} j_m! }{i_{m-1}!j_{m-1}!} (-1)^{L^{(q)}_{m}+i_{m}+1} \ \times \\
&\ \ \ \sum_{\alpha_{m}=0}^{L^{(q_{m+1})}_{m}+i_{m}} \frac{ (\alpha_{m}+j_{m})! }{\alpha_{m}!j_{m}!}
\frac{ \partial_{s_{m-1}}^{\alpha_{m}+j_{m}-i_{m-1}-j_{m-1}}}{(\alpha_{m}+j_{m}-i_{m-1}-j_{m-1})!}
\frac{ \partial_{s_{m}}^{L^{(q_{m+1})}_{m}+i_{m}-\alpha_{m}}}{(L^{(q_{m+1})}_{m}+i_{m}-\alpha_{m})!}
\Gamma[c^{(q_{m+1})}-s_{m}]{\rm RHS}_{i_mj_m}\bigg|_{s_{m}=0}\notag
\end{align}
where we defined, 
\begin{equation}
c^{(0)}=1\quad;\quad c^{(\mathbb{N})}=\mathbb{N}\qquad;\qquad L_m^{(0)}=L_n\quad;\quad L_{m}^{(\mathbb{N})}=L_m-1,
\end{equation}
\end{mdframed}
Going from \eqref{ini_rec} to \eqref{rec_for0} we did the following:
\begin{itemize}
\item[A)] 
We {\tt ibp} the power of $s_m$ in $\frac{ \Gamma[1-s_{m}] }{(-s_{m})^{L_{m}+1+i_{m}}  }$ for $q_{m+1}=0$ and 
$\frac{\Gamma[p_{m+1}-s_{m}] }{(-s_{m})^{L_{m}+i_{m} }  }$  for $p_{m+1}=\mathbb{N}$. This means using the manipulation:
\begin{equation}
\frac{X}{(-s)^{\ell+1}}=-\frac{X}{\ell!}\,\partial_s^{\ell} \bigg(\frac{1}{s}\bigg)\ \underbracket{\ = \ }_{{\tt ibp}}\ \frac{ (-)^{\ell+1}}{\ell!}\frac{\partial_s^\ell X}{s} 
\end{equation} 
valid for any $X$.
\item[B)]
We distributed the derivatives according to  $\alpha_m+\beta_m=L_m^{(q_{m+1})}+i_m$ 
where $\alpha_m$ added to $\partial_{s_{m}}^{ {\alpha_m+ j_m} }\Gamma[s_{m}-s_{m-1}]$. 
The residue of the latter in $s_m=0$ is found by switching the derivatives, so to get 
$(-1)^{\alpha_m+j_m} \partial^{\alpha_m+j_m}_{s_{m-1}}\Gamma[-s_{m-1}]$. 
\item[C)]
We {\tt ibp} these $(-1)^{\alpha_m+j_m} \partial^{\alpha_m+j_m}$ derivatives in order 
to produce $\frac{ \Gamma[-s_{m-1}]}{(-s_{m-1})^{L_{m-1}+i_{m-1}}}$.
\end{itemize}
In doing C) we broke $\prod_{k=2}^{m-1}\frac{ \Gamma[s_k-s_{k-1}]}{(-s_k)^{L_k}}=\prod_{k=2}^{m-2}\frac{ \Gamma[s_k-s_{k-1}]}{(-s_k)^{L_k}}\frac{ \Gamma[s_{m-1}-s_{m-2}]}{(-s_{m-1})^{L_{m-1}}}$ and distributed the 
derivatives according to $i_{m-1}+j_{m-1}+k_{m-1}=\alpha_m+j_m$. 
To finalize, we interchanged the two sums using, 
\begin{equation}
\sum_{\alpha_{m}=0}^{\ell+i_{m}} \sum_{i_{m-1}+j_{m-1}=0}^{\alpha_{m}+j_{m}}=
\sum_{i_{m-1}+j_{m-1}=0}^{\ell+i_{m}+j_{m}}\ \sum_{\alpha_{m}=\max(0,i_{m-1}+j_{m-1}-j_{m})}^{\ell+i_{m}}
\end{equation}
Note that the max can be dropped thanks to a factorial whose
argument is $(L^{(q_{m+1})}_{m}+i_{m}-\alpha_{m})$. Thus, after C), we have found an integrand as in \eqref{ini_rec} but now for $s_{m-1}$.\\

The recursive step in \eqref{inprec} --
generated by considering the residue in $s_m=p_{m}=\mathbb{N}$ -- gives
\begin{mdframed}
\begin{align}\label{rec_forp}
&
{\rm RHS}_{i_{m-1}j_{m-1}}^{\ q_{m+1}\rightarrow p_m}\Big[i_m,j_m,{\rm RHS}_{i_mj_m}\Big]= 
\frac{ (L_{m-1})_{i_{m-1}} j_m!}{i_{m-1}!j_{m-1}!}
\frac{ \partial_{s_{m-1}}^{j_m-i_{m-1}-j_{m-1}}}{(j_m-i_{m-1}-j_{m-1})!}
\frac{(-)^{p_m-q_{m+1}-1}}{(p_m-q_{m+1})!} \frac{{\rm RHS}_{i_mj_m}}{(-s_m)^{L_m+i_m}}\bigg|_{s_{m}=p_n\ge 1}  
\end{align}
\end{mdframed}
Note that $p_m\ge 1$, always. However, when $q_{m+1}\ge 1$ 
it follows that $p_m\ge p_{m+1}$, and a structure of Young diagrams is generated 
from summing over  $p_{m+1}$ and $p_m$. This is automatically enforced by
 the factorial whose argument is $(p_m-q_{n+1})$. Going from \eqref{ini_rec} to 
 \eqref{rec_forp} we only have to take the residue of $\Gamma[q_{m+1}-s_{m}]$, 
 for any value of $q_{m+1}$, and  {\tt ibp} $(-1)^{j_m} \partial^{j_m}_{s_{m-1}}\Gamma[p_{m}-s_{m-1}]$.\\

Running the above recursion up to $m\ge2$ included returns a set of form $\{ (i_1,j_1), q_2,{\rm RHS}\}$ which will be attached to the $s_1$ integration, as follows,
\begin{align}
I_{i_1j_1}^{\,q_2}= \bigg[\partial_{s_1}^{j_1} \frac{1}{\Gamma[1-s_1]}\bigg]
\frac{\Gamma[q_2-s_1]  }{ \ \ (-s_1)^{L_1+i_1}}\,  {\rm RHS}^{\,q_2}_{i_1j_1}
\end{align}
The structure of this final integrand is again dictated by the integers $i_1,j_1$ 
and the value of $q_2$, ie.~the \emph{data}. We finish the routine with the instructions:
\begin{mdframed}
\begin{align}
{\rm RHS}^{\ q_2\rightarrow 0}[i_1,j_1,{\rm RHS}_{i_1j_1}]=&\ (-)^{L_1^{(q_2)}+i_1}
\sum_{\alpha_1=0}^{L_1^{(q_2)}+i_1 }
\bigg[\frac{\partial_{s_1}^{\alpha_1+j_1}}{\alpha_1!}\Gamma[1-s_1]^{-1}\bigg]
\bigg[\frac{ \partial_{s_1}^{L_1^{(q_2)}+i_1-\alpha_1} }{  (L_1^{(q_2)}+i_1-\alpha_1)!}\Gamma[c^{(q_2)}-s_1] {\rm RHS}_{i_1j_1}\bigg]\bigg|_{s_1=0} \\
{\rm RHS}^{\ q_2\rightarrow p_1}[i_1,j_1,{\rm RHS}_{i_1j_1}]=&\ \bigg[ \sum_{k=1,3,5,\ldots}^{j_1}\!\!\!(-)^{p_1} \frac{j_1!  }{(k)!(j_1-k)! }(i\pi)^{k-1}\Gamma^{(j_1-k)}[p_1]\bigg]\frac{ (-)^{p_1-q_2-1} }{(p_1-q_2)! } \frac{ {\rm RHS}_{i_1j_1}}{ (-s_1)^{L_1+i_1}}\bigg|_{s_1=p_1\ge1}
\end{align}
\end{mdframed}
In the last formula we used reflection $\Gamma[1-z]^{-1}=\frac{\sin(\pi z)}{\pi}\Gamma[z]$. In the same 
formula, if $j_1=0$, then correctly there is no residue because $\Gamma[q_2-s_1]/\Gamma[1-s_1]$ is regular at integers.
\\

In all cases, the derivative structure of the various RHS can be expanded with a 
computer, ie.~{\tt Mathematica}. It is interesting 
to note that all over at intermediate steps 
there are {\tt EulerGamma} contributions, but these cancel out.
The case of the sequence $\{s_{j}=0\ \forall j\ge 1\}$ is perhaps the most 
straightforward case to check. For illustration, we quote some cases at 5pt, namely,
\begin{align}
L_1=L_2=1\qquad;\quad
{\rm RHS}^{0\rightarrow0\rightarrow 0}=& 
-\tfrac{\pi^2}{6}\widetilde{\cal T}(\underline{0})+\widetilde{\cal T}^{(1,1)}(\underline{0})+\tfrac{1}{2}\widetilde{\cal T}^{(2,0)}(\underline{0}) \\[.2cm]
L_1=L_2=2\qquad;\quad
{\rm RHS}^{0\rightarrow0\rightarrow 0}=& 
+\tfrac{7\pi^4}{360}\widetilde{\cal T}(\underline{0})-\zeta_3 \widetilde{\cal T}^{(0,1)}(\underline{0}) 
-\tfrac{\pi^2}{6}\widetilde{\cal T}^{(1,1)}(\underline{0})+\tfrac{1}{4}\widetilde{\cal T}^{(2,2)}(\underline{0})+
+\tfrac{1}{6}\widetilde{\cal T}^{(3,1)}(\underline{0})+\tfrac{1}{24}\widetilde{\cal T}^{(4,0)}(\underline{0})  \notag
\end{align} 
and a case at 6pt, namely
\begin{align}
L_{i=1,2,3}=1\qquad;\quad {\rm RHS}^{0\ldots 0}= 
-\tfrac{\pi^2}{6}( \widetilde{\cal T}^{(1,0,0)}(\underline{0}) +{\rm cyclic}) + \widetilde{\cal T}^{(1,1,1)}(\underline{0})+\tfrac{1}{2}
( \widetilde{\cal T}^{(2,1,0)}(\underline{0}) +{\rm cyclic}) +\tfrac{1}{6}  \widetilde{\cal T}^{(3,0,0)}(\underline{0})
\end{align}
It is straightforward to check the maxlog in \eqref{Tas_string}.
Beyond the maxlog formula, the knowledge of the $R$-matrix is crucial in many ways, 
even in the examples above, it is needed to compute derivatives of ${\cal T}$.

%
%

\section{Worked example: 5-pt split-ladders in $(--)$ signature}

In this section we discuss the integration of the 5pt split-ladders ${\cal L}_{L_1,L_2}$. The idea is to exemplify
the recursion relations derived in the previous section with a concrete example. In addition, we will make explicit the stratification in $\log$s.
The case $L_1=L_2=1$ was studied by Drummond and Druc in the PhD thesis \cite{tesi_James}. 
Borrowing their results we were able to check independently that our representation for the integral ${\cal I}_{1,1}$ satisfies the known differential equation, 
and matches the series expansion of the symbol \cite{James_symbol}. \\

The 5-pt integrand is
 \begin{equation} 
{\cal L}_{L_1L_2}=
\frac{\rho_1^{s_1}\rho_2^{s_2} }{ \Gamma[1-s_1]}
\frac{  \Gamma[s_2-s_1]\Gamma[-s_2]  }{\ \ \  (-s_1)^{L_1} \, (-s_2)^{L_2}} \times \frac{ {\cal T}_{a_1a_2}(s_1,s_2)}{(a_1+s_1)^{L_1}(a_2+s_2)^{L_2}}\,,
\end{equation}
%
%

We will integrate first in $s_2$ and then in $s_1$. 
Of the four possible sequences $s_j\in\{0,\mathbb{N}\}$ only three are realised since 
the sequence $\{p_1,p_2\}$ does not have a residue. In the order we will consider taking residues 
\begin{align}
{\cal L}_{L_1L_2}\Bigg|_{s_2=0,s_1=0}
\qquad;\qquad 
{\cal L}_{L_1L_2}\Bigg|_{s_2=0,s_1=\mathbb{N}}
\qquad;\qquad
{\cal L}_{L_1L_2}\Bigg|_{s_2=\mathbb{N},s_1=0}
\end{align}
The sum of these three gives the integral ${\cal I}_{L_1,L_2}$. We understand above the sum over $\underline{a}$.\\

Let us start by taking the residue of the $s_2=0$ pole we find
\begin{align}
{\cal L}_{L_1L_2}\Bigg|_{s_2=0}\ \ 
&\ =\ (-)^{L_2+1} 
\sum_{i_2=0}^{L_2}\sum_{i_1=0}^{L_2-i_2} \frac{(\log \rho_2)^{i_2}}{i_2!}
\frac{\rho_1^{s_1}}{(-s_1)^{L_1+1}}\frac{ (-1)^{i_1}\Gamma^{(i_1)}[-s_1]}{i_1!\Gamma[-s_1]} \,{\cal C}_{L_2-i_1-i_2} \label{5ptfors1}
\end{align}
where ${\cal C}_{M}\equiv\frac{1}{M!} \partial^{M}_{s_2}\bigg[ \Gamma[1-s_2]\,\frac{ {\cal T}_{a_1a_2}(s_1,s_2)}{(a_1+s_1)^{L_1}(a_2+s_2)^{L_2}}\bigg]_{s_2=0}$.
Then
\begin{align}
{\cal L}_{L_1L_2}\Bigg|_{s_2=0,s_1=0}
&\!\!\!
=\frac{\ \ \ \ (-)^{L_1+L_2}}{(L_1-1)!} \sum_{i_2=0}^{L_2}\sum_{i_1=0}^{L_2-i_2}\sum_{j_1=0}^{L_1+L_2-i_1-i_2} 
\frac{(\log \rho_2)^{i_2} (\log\rho_1)^{i_1+j_1} }{i_1!i_2!j_1!} \times  C^{(0,0)}_{i_1+i_2,j_1}
\end{align}
where we acted with the $i_1$ derivatives from {\tt ibp} and then we manipulate the various sums in order to find
\begin{equation}
C^{(0,0)}_{I_{12},j_1}=\sum_{j_2=\max(0,j_1-L_1)}^{L_2-I_{12}}\!\!\!\!\frac{(-1)^{j_2} (j_2+L_1-1)!}{j_2!(j_2-j_1+L_1)!} \ \partial_{s_1}^{L_1-j_1+j_2}\bigg[ \Gamma[1-s_1]
\sum_{j_3=0}^{L_2-I_{12}-j_2} \frac{1}{j_3!} \partial^{j_3}_{s_1} \bigg[ \frac{ {\cal C}_{L_2-I_{12}-j_2-j_3} }{\Gamma[1-s_1]} \bigg]\bigg]_{s_1=0}
\end{equation}
As pointed out already, the max in the summation can be set to zero because of  $(j_2-j_1+L_1)!$ -- a sign of analyticity.
Similarly
\begin{align}
{\cal L}_{L_1L_2}\Bigg|_{s_2=0,s_1=p_1}\ \ 
&\!\!\!
=(-)^{L_2+1}\sum_{i_2=0}^{L_2}\sum_{i_1=0}^{L_2-i_2} \sum_{j_1=0}^{L_2-1-i_1-i_2} \frac{(\log \rho_2)^{i_2} (\log\rho_1)^{i_1}}{i_1!i_2!} 
 C^{(p_1,0)}_{i_1+i_2,j_1}
 \end{align}
where
\begin{equation}
C^{(p,0)}_{I_{12},j_1}= \bigg[ \sum_{k=0,2,\ldots}^{j_1} \frac{ (i\pi)^{k}\Gamma^{(j_1-k)}[p+1] }{(k+1)!(j_1-k)! p!} \ \ \bigg]
\,\rho_1^p\,\times\!\!\sum_{j_2=0}^{L_2-1-I_{12}-j_1}\frac{1}{j_2!} \partial_{s_1}^{j_2}
\bigg[ \frac{  {\cal C}_{L_2-1-I_{12}-j_1-j_2} }{(-s_1)^{L_1+1} }\bigg]_{s_1=p}
\end{equation} 
This contribution will have to be further summed over $p_1\in\mathbb{N}$. \\

The last sequence to consider is 
\begin{align}
&
{\cal L}_{L_1L_2}\Bigg|_{s_2=p_2,s_1=0}= \\ 
&(-1)^{L_1}\sum_{i_1=0}^{L_1-1} \frac{(\log\rho_1)^{i_1} }{i_1!} \bigg[ \frac{(-)^{p_2-1} }{(-p_2)^{L_2}p_2!} \bigg]\rho_2^{p} \times \frac{ \partial_{s_1}^{L_1-1-i_1} }{(L_1-1-i_1)!}\bigg[  \frac{ \Gamma[p_2-s_1] }{\Gamma[1-s_1]}
\frac{ {\cal T}_{a_1a_2}(s_1,s_2)}{(a_1+s_1)^{L_1}(a_2+s_2)^{L_2}}\bigg]_{s_2=p_2,s_1=0} \notag
\end{align}
All these formulae are quite straightforward to implement in a computer.\\

We will now look at the specific example of $L_1=L_2=1$. The stratification in $\log\rho_{i=1,2}$ reads
\begin{align}
{\cal I}_{1,1}=&\sum_{a,b\ge 1}\bigg[ (\log\rho_2)\Big[\, {\cal T}_{ab}(\underline{0})  (\log\rho_1)  + {\cal T}_{ab}^{(1,0)}(\underline{0})  \Big] \\
&\rule{2.2cm}{0pt}\Big[ \tfrac{1}{2} {\cal T}_{ab}(\underline{0}) (\log\rho_1)^2 + 
\big( {\cal T}^{(1,0)}_{ab}(\underline{0})+{\cal T}^{(0,1)}_{ab}(\underline{0}) \big) (\log\rho_1)\, + \\[.2cm]
&\rule{2.2cm}{0pt} 
\Big( \tfrac{1}{2}{\cal T}^{(2,0)}_{ab}(\underline{0}) +{\cal T}^{(1,1)}_{ab}(\underline{0}) -\tfrac{\pi^2}{6} {\cal T}_{ab}(\underline{0}) 
+\sum_{p\ge 1} (-1)^p\, \frac{\rho_1^p {\cal T}_{ab}(p,0) -\rho_2^p {\cal T}_{ab}(0,p) }{ p^2} \Big) \bigg]
\end{align}
where the ``diagonal line" of maxlogs is $\big((\log\rho_2)(\log\rho_1) + \frac{1}{2}(\log\rho_1)^2 \big) {\cal T}_{ab}(\underline{0})$.
Following \cite{tesi_James}, ${\cal L}_{1,1}$ is ``boxable'' on the external points, 
\begin{equation}\label{boxable}
\Box_i\frac{1}{x^2_{ij}}=-4\pi^2 \delta^4(x_i-x_j)
\end{equation}
and in particular it simplifies to a 4pt one-loop ladder when acted upon with \eqref{boxable}.
By combining $\Box_1,\Box_2,\Box_{1+2}$, a differential equation can be derived \emph{on the plane} \cite{tesi_James}. 
In our notation  it reads
\begin{align}\label{diff_equaZ}
\Big[\rho_2 \mathbb{O}(\overline{\mathbb{O}}+1)\rho_1\Big]{\cal I}_{1,1}= 
+ (Z_1 Z_2-\bar{Z}_1 \bar{Z}_2) R_1
+ \frac{ (Z_2-\bar{Z}_2) (Z_2 Z_1(1-\bar{Z}_1) - \bar{Z}_2 \bar{Z}_1(1-{Z}_1) )}{ Z_2(1-\bar{Z}_1)- \bar{Z}_2(1-{Z}_1)-Z_2\bar{Z}_2 (Z-\bar{Z}_1) } R_2
\end{align}
[[to compare with \cite{tesi_James} $a_{{\rm their}, 1}=(Z_1-1)/(Z_1Z_2), \,a_{{\rm their},2}=Z_2^{-1}$ ]] where
\begin{align}
\mathbb{O}=&\ (Z_2-\bar{Z_2})\partial_{Z_2} +(Z_1-\bar{Z}_1)\partial_{Z_1}+\frac{(Z_1-\bar{Z}_1)^2}{Z_1 \bar{Z}_1} Z_1\partial_{Z_1} \quad;\quad
R_1=\ \phi(Z_2)-\phi(\tfrac{Z_1 Z_2}{Z_1-1})\quad,\quad R_2=\phi(\tfrac{ Z_2}{(Z_1-1)(1-Z_2)})
\end{align}
and $\phi(x)=-(\log x+\log\bar{x})\big( {\rm Li}_1(x)-{\rm Li}_1(\bar{x}) \big)+2 \big( {\rm Li}_2(x)-{\rm Li}_2(\bar{x}) \big)$ is 
the pure part of the box integral \cite{Usyukina:1992jd,Isaev:2003tk}. Interestingly, 
the rhs of \eqref{diff_equaZ} does not have a $\log\rho_1\log\rho_2$ coefficient 
function, neither a $(\log\rho_1)^2$. Thus the differential operator has to annihilate 
the maxlogs coefficient functions. Indeed, it annihilates all trace polynomials evaluated in zero, namely
\begin{equation}
\Big[\rho_2 \mathbb{O}(\overline{\mathbb{O}}+1)\rho_1\Big] {\cal T}_{a_1a_2}(\underline{0})=0\qquad;\qquad
{\cal T}_{a_1a_2}(\underline{0})= 
\frac{ (-\bar{Z}_2 \bar{Z}_1\partial_{\bar{Z}_1} - Z_2Z_1\partial_{Z_1} )^{a_2-1} }
{(a_2-1)!} \frac{Z_1^{a_1}-\bar{Z}_1^{a_1}}{Z_1-\bar{Z}_1}
\end{equation}
and therefore it annihilates the maxlog $\mathbb{F}^{(--)}_{k_1k_2}$, which is a linear combination of them. In particular,
\begin{equation}
\mathbb{F}^{(--)}_{k_1k_2}=\frac{(-1)^{L_1+L_2}}{L_1!L_2!k_1!k_2!} 
\sum_{\substack{a\in\mathbb{N}\\b\in\mathbb{N}}}\frac{(-1)^{b-1}}{a^{L_1} b^{L_2}}\sum_{i,j\ge0}\Big[\,^{i}_{\rule{0pt}{.25cm}j}\Big]\Big[\,^{a-1-i}_{\rule{0pt}{.25cm}b-1-j}\Big]
Z_1^iZ_2^j\bar{Z}_1^{a-1-i}\bar{Z}_2^{b-1-j}
\end{equation}
where $k_1+k_2=L_1+L_2$.


\subsection{MPLs for the maxlog of ${\cal L}_{1,1}$}

From \cite{tesi_James}, we can check explicitly our series representation for ${\cal L}_{1,1}$. We did this in general, finding perfect agreement. 
Here for simplicity we will give some more details about the maxlog,
which in particular reads
\begin{equation}
\mathbb{F}^{(--)}_{11}=\sum_{a_2\in\mathbb{N}}
\frac{ \mathcal{D}^{(a_2)}_{2,1}}{a_{2}}\left[ \frac{ {\rm Li}_{1}(W_1) - {\rm Li}_{1}(\bar{W}_1) }{W_1-\bar{W}_1}\right]
\end{equation}
Integrating the symbol in \cite{tesi_James}, in terms of polylogs, the result is
\begin{align}
(W_2-\bar{W}_2)\sum_{a_2\in\mathbb{N}}
\frac{ \mathcal{D}^{(a_2)}_{2,1}}{a_{2}}\left[ \frac{ {\rm Li}_{1}(W_1) - {\rm Li}_{1}(\bar{W}_1) }{W_1-\bar{W}_1}\right]=\ &
+ (G[1, W_1] -G[1, \bar{W}_1] ) \ {\rm Li}_1\big({W_2-\bar{W}_2}\big/{W_1-\bar{W}_1}\big)\notag\\[-.35cm]
&
+  G[W_1 - \bar{W}_1, -1 + W_1, W_2]  \notag \\
&
-  G[-W_1 + \bar{W}_1 + W_2, -1 + \bar{W}_1, \bar{W}_2]  \notag \\
&
+   G[\tfrac{-1 + \bar{W}_1}{-1 + W_1}W_2, -1 + \bar{W}_1, \bar{W}_2] \notag\\
&
-  G[0, -1 + \bar{W}_1, \bar{W}_2] \notag \\
&
+  G[-1 + W_1, W_2] G[-W_1 + \bar{W}_1 + W_2, \bar{W}_2]   \notag \\
&
- G[-1 + W_1, W_2]G[\tfrac{-1 + \bar{W}_1 }{-1 + W_1}W_2, \bar{W}_2]  
\label{polylogs_Omer}
\end{align}
Note that ${\rm Li}_1\big(\tfrac{W_2-\bar{W}_2}{W_1-\bar{W}_1}\big)=  G[+W_1 - \bar{W}_1, W_2] + G[-W_1 + \bar{W}_1 + W_2, \bar{W}_2]$
comes straightforwardly from $\mathcal{D}^{(a_2)}_{2,1}[ \frac{1}{(W_1-\bar{W}_1)}]$.~\\

Let us observe that the series on the l.h.s.~of \eqref{polylogs_Omer}  is manifestly symmetric 
under the exchange $W_{i=1,2}\leftrightarrow \bar{W}_{i=1,2}$, 
but the r.h.s.~of \eqref{polylogs_Omer} is not. Obviously, both agree as we series expand. 
To check this we can use the simple formulae, 
\begin{align}
G[\overbrace{0,\ldots,0}^{n},A;X]=&-{\rm Li}_{n+1}(\tfrac{X}{A})\\[.2cm]
G[A\,,B;X]=&+{\rm Li}_2(\tfrac{A}{A-B})-{\rm Li}_2(\tfrac{A-X}{A-B}) +\log(1-\tfrac{A}{B})\log(1-\tfrac{X}{A}) \notag
\end{align}
Now, there is a trivial way of making manifest the symmetry, add and subtract.  
We know then that the antisymmetric term has to vanish. This is a non trivial identity, 
and can be checked with the help of ${\rm Li}_2$ identities \cite{DZagier}.

\section{More on the 5-pt split-ladders}

As mentioned in the main body of our letter, there is a variety of signatures one could consider. 
So far we focussed on $(-,\ldots,-)$, since gave us a nice recursion wrt the number of points, and since
we expect that by knowing the expansion of the coefficient functions of the log stratification \eqref{log_strata} in a 
given signature provides enough data to resum the diagram in terms of pure functions. 
But other signatures can be equally interesting. 
In this section we describe the example of the 5-pt split ladders in  
$(+-)$ signature.  This corresponds to crossing $Z_1\rightarrow Z_1^{-1}$, 
and allows us to check a non trivial relations w.r.t.~the $(--)$ representation. 
As a byproduct, our computations provide a further consistency check of the SoV representation itself.\\

\subsection{$(+-)$ signature}

From ${\cal F}_{L_1,L_2}$ in \eqref{calFdef} we consider
\begin{equation}
{\bf i}\, u_1= -\tfrac{a_1}{2}-t_1\qquad;\qquad {\bf i}\, u_2= \tfrac{a_2}{2}+t_2\,,
\end{equation}
and for convenience of comparison we will redefine $Z_1^{-1}=z_1$, $Z_2=z_2$. Similarly $\tilde\rho_i=z_i\bar{z}_i$. The integrand now reads
\begin{equation}
\widetilde{\cal L}_{L_1,L_2}=
(\tilde\rho_1)^{t_1}(\tilde\rho_2)^{t_2} 
\frac{  \ \Gamma[-t_1]\Gamma[-t_2]  }{\ \ \  (-t_1)^{L_1} \Gamma[1-t_1-t_2](-t_2)^{L_2}} \times \widetilde{\cal T}_{a_1a_2}(t_1,t_2)
\end{equation}
where 
\begin{equation}
\widetilde{\cal T}_{ \underline{a} }=\frac{1}{(a_1+t_1)^{L_1}(a_2+t_2)^{L_2}}
\frac{a_1 a_2 }{ \frac{ \Gamma[1+a+t_1] \Gamma[1+b+t_2] }{ \Gamma[a_1+t_1+t_2]}}\times z_1 \bar{z}_1\sum_{i=0}^{a-1}\sum_{j=0}^{a_2-1}  r_{a_1i,a_2,a_2-j-1}(-t_{1}-t_2-a)\  z_1^{i} z_2^{j} \bar{z}_1^{a_1-1-i} \bar{z}_2^{a_2-1-j}
\end{equation}
%
%
\begin{figure*}[b]
\centering
\includegraphics[scale=0.65]{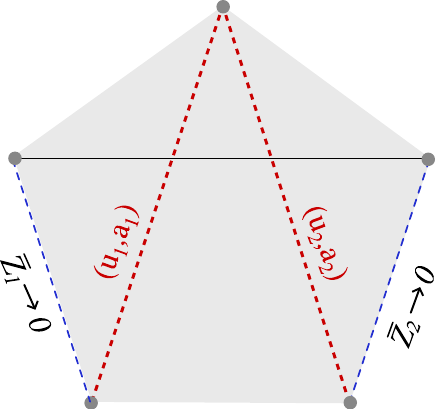}
\caption{The split-ladder Feynman diagram at $5$-pts. 
The blue lines represent the kinematic limit of the $(+-)$ signature, to be contrasted with FIG.\ref{slicing} (left) for the $(--)$ signature.}
\label{5_kin}
\end{figure*}
The various building blocks are of course the same as for ${\cal L}_{L_1,L_2}$, however there are some small but important differences.
To start with, the $(+-)$ signature appears be produce a symmetric integrand:  With reference to FIG.\ref{5_kin},
this symmetry is due to the fact that the small $z_1,z_2$ expansion is the expansion about the 
central segment on the bottom of the pentagon made by the external points.
-- More generally, this configuration is possible for external points $n$ odd. --
Second, note the change  $\Gamma[s_2-s_1]^{+1}$ vs $\Gamma[1-t_1-t_2]^{-1}$ 
when comparing $(--)$ and $(+-)$ signature. This changes demonstrates 
in formulae that the kinematics of the small $z_1,z_2$ expansion will not be sequential.\\

Proceeding with the integration we have three sequences: $(s_1=0,s_2=0)$, then $(p_1,0)$ and $(0,p_2)$. We find
\begin{align}
\widetilde{\cal L}_{L_1L_2}\Bigg|_{t_2=0,t_1=0}
&\!\!\!
= (-)^{L_1+L_2} \sum_{i_2=0}^{L_2}\sum_{i_1=0}^{L_1}
\frac{(\log \tilde\rho_2)^{i_2} (\log\tilde\rho_1)^{i_1} }{i_1!i_2!(L_1-i_1)!(L_2-i_2)!} \times  
\partial_{t_1}^{L_1-i_1}\bigg[ \Gamma[1-t_1]\partial^{L_2-i_2}_{t_2}\bigg[
\frac{\Gamma[1-t_2]\, \widetilde{\cal T}_{ \underline{a} }}{\Gamma[1-t_1-t_2]}\bigg]\bigg]_{\substack{t_1=0\\t_2=0}}
\end{align}
\begin{align}
\widetilde{\cal L}_{L_1L_2}\Bigg|_{t_2=0,t_1=p_1}
&\!\!\!
= (-)^{L_2} \sum_{i_2=0}^{L_2}\sum_{j_2=0}^{L_2-i_2}
\frac{(\log \tilde\rho_2)^{i_2} (\tilde\rho_1)^{p_1} }{i_2!(L_2-i_2-j_2)!} \times \!\!\!\! 
\sum_{k=1,3,\ldots} \frac{ (i\pi)^{k-1} \Gamma^{(j_2-k) }(p_1) }{(j_2-k)!k! (-p)^{L_1}p!}
\partial^{L_2-i_2-j_2}_{t_2}\bigg[
\Gamma[1-t_2]\, \widetilde{\cal T}_{ \underline{a} } \bigg]_{\substack{t_1=p_1\\t_2=0}}
\end{align}
\begin{align}
\widetilde{\cal L}_{L_1L_2}\Bigg|_{t_2=p_2,t_1=0}
&\!\!\!
= (-)^{L_1} \sum_{i_1=0}^{L_1}\sum_{j_1=0}^{L_1-i_1}
\frac{(\log \tilde\rho_1)^{i_1} (\tilde\rho_2)^{p_2} }{i_1!(L_1-i_1-j_1)!} \times \!\!\!\! 
\sum_{k=1,3,\ldots} \frac{ (i\pi)^{k-1} \Gamma^{(j_1-k) }(p_2) }{(j_1-k)!k! (-p)^{L_2}p!}
\partial^{L_1-i_1-j_1}_{t_1}\bigg[
\Gamma[1-t_1]\, \widetilde{\cal T}_{ \underline{a} } \bigg]_{\substack{t_1=0\\t_2=p_2}}
\end{align}
Of course, the last two expressions are symmetric under $1\leftrightarrow 2$.

\subsection{Crossing from $+$ to $-$}

From resumming ${\cal I}_{L_1,L_2}$ in both signatures, $(--)$ and $(+-)$, we can see that the transformation
$z_1\rightarrow Z_1$ and $\bar{z}_1\rightarrow \bar{Z}_1$ exchanges the two. 
This check can be done at a simplified level,  by looking
at a given power of $\log(Z_2{\bar Z}_2)=\log(z_2{\bar z}_2)$, and inspecting the 
corresponding coefficient functions.  
We give some details in the following.~\\ 

For simplicity, we will look at contributions of the form
$\log^{L_2}(\rho_2)\times Z_2^{N-n} {\bar Z}_2^n\times m_{N,n}(Z_1,\bar{Z_1})$ for fixed values of $N,n$.
Depending on the signatures we will label it with $m^{\pm}$.
The general form is
\begin{align}
m_{N,n}(Z_1,\bar{Z_1})=
\sum_{\ell=0}^{L_1} \log^{\ell_1}(\rho_1)\Bigg[ 
\frac{ p_{\ell}(Z_1,{\bar Z}_1)  {\rm Li}_{2L_1-\ell}(Z)+  q_{\ell}(Z_1,{\bar Z}_1){\rm Li}_{2L_1-\ell}({\bar Z}_1)  }{(Z_1-{\bar Z}_1)^{N+1}  }+\ldots \Bigg]
\end{align}
For example, when $N=3,n=1$ with $L_1=L_2=2$, we find that $m^-_{N,n}(Z_1,\bar{Z_1})$ contains the following contributions,
\begin{align}
&
\ell=0\quad;\quad
-\tfrac{9Z_1^2\bar{Z}_1}{16}\tfrac{{\rm Li}_4(Z_1)-{\rm Li}_4(\bar{Z}_1)}{ (Z_1-\bar{Z}_1)^4 }
+\tfrac{3(2Z_1^2+5 Z\bar{Z_1}-\bar{Z}_1^2)}{64}\tfrac{{\rm Li}_3(Z_1)-{\rm Li}_3(\bar{Z}_1)}{(Z_1-\bar{Z}_1)^3}
-\tfrac{10Z_1+\bar{Z}_1}{192}\tfrac{{\rm Li}_2(Z_1)-{\rm Li}_2(\bar{Z}_1)}{(Z_1-\bar{Z}_1)^2}
+\tfrac{1}{192}\tfrac{{\rm Li}_1(Z_1)+{\rm Li}_1(\bar{Z}_1)}{(Z_1-\bar{Z}_1)}\notag\\[.2cm]
&
\ell=1\quad;\quad \\
&
\!\!\!+\tfrac{9Z_1^2\bar{Z}_1}{32}\tfrac{{\rm Li}_3(Z_1)-{\rm Li}_3(\bar{Z}_1)}{(Z_1-\bar{Z}_1)^4}
-\tfrac{(Z_1^2+5Z_1\bar{Z}_1-\bar{Z}_1^2){\rm Li}_2(Z_1)}{32(Z_1-\bar{Z}_1)^3}
-\tfrac{(4Z_1^2+5Z_1\bar{Z}_1-\bar{Z}_1^2){\rm Li}_2(\bar{Z}_1)}{64(Z_1-\bar{Z}_1)^3}
+\tfrac{(5Z+2\bar{Z}) {\rm Li}_1(Z)-(5Z+\bar{Z}){\rm Li}_1(\bar{Z})}{192(Z-\bar{Z})^2}
-\tfrac{Z_1}{192(1-Z_1)(Z_1-\bar{Z}_1)}\notag \\[.2cm]
&\ell=2\quad;\quad 
-\tfrac{3Z_1^2\bar{Z}_1}{64}\tfrac{ {\rm Li}_2(Z_1)-{\rm Li}_2(\bar{Z}_1)}{(Z_1-\bar{Z}_1)^4}
-\tfrac{\bar{Z}_1(\bar{Z}_1-5Z_1)}{128}\tfrac{ {\rm Li}_1(Z_1)}{(Z_1-\bar{Z}_1)^3}+\tfrac{Z_1^2}{64}\tfrac{ {\rm Li}_1(\bar{Z}_1) }{(Z_1-\bar{Z}_1)^3}
-\tfrac{Z_1\bar{Z}_1}{(1-Z_1)(Z_1-\bar{Z}_1)^2}\notag
\end{align}
Upon $Z_1=\frac{1}{z_1}$, using identities for the polylogs such that we can expand in small $z_1$,
we match $m^+_{N,n}(Z_1,\bar{Z_1})$.\\

One feature of the diagram that we would like to emphasise here is that generically 
the maximal power of $\log (z_1\bar{z}_1)$ is $L_1$, while the maximal power 
of  $\log(Z_1\bar{Z}_1)$ is $L_1+L_2$.  This means that the arguments of the 
polylogs in the resummation include the variable $Z_1-\bar{Z}_1$ etc..., so 
that upon $Z_1=\frac{1}{z_1}$ the $(--)$ signature matches the $(+-)$ signature.

\subsection{Decoupling effect: kinematics and integral prescription}

In order to define rigorously the SoV integral we need to pay attention to the contour of integration.
Indeed, no matter what signature of integration is chosen, 
once we restrict ourselves to two subsequent cuts with excitations $(v,b)$ 
and $(u,a)$ there is a pole  when $a=b$ on the integration contour $u=v$, 
due to the $\Gamma$ function 
\begin{equation}
\Gamma\left(-{\bf i}u+{\bf i}v +\frac{a-b}{2}\right)\,
\end{equation}
that enters $H_{a,b}(u,v)$.
This type of pole is called \emph{decoupling pole} because it identifies excitation $(u,a)$ and $(v,b)$. 
In the picture where the SoV integrand represents the passage of excitations from one cut to the other, this pole
is effectively decoupling their interaction. Let's see the consequences of this. \\

At the point $u=v$, $a=b$ the following holds 
\begin{equation}
\tilde R_{a,b}(0) = \mathbb{P}_{a,b} \in \text{End}(V_a\otimes V_b) \qquad;\qquad \text{Res}\Big|_{\substack{v=u \\ a=b}} \left( \tilde H_{a,b}(u,v) \right) = E_a(u) \frac{(-1)^{a-1}}{a}\,,
\end{equation}
where $\mathbb{P}_{a,b}$ is the permutation of factors in the tensor product.  
In particular, the integrand with two excitations simplifies as follows, 
\begin{align}
\begin{aligned}
\label{the_decoupling}
\text{Res}_{\substack{v=u \\ a=b}} \left[ \,E_{a}(u)^{L_1+1} E_{b}(v)^{L_2+1} \tilde H_{a,b}(u,v) \,\text{Tr}_{b} \left[
\left(\frac{Z_1}{\bar{Z}_1}\right)^{{J}_{3;a }} \!\!\!\!\!\!\! \otimes \left(\frac{Z_2}{\bar{Z}_2}\right)^{{J}_{3;b }} \,\,\cdot \,
\tilde{R}_{a,b}(u,v) \right]\right] =  \frac{1}{a} E_{a}(u)^{L_1+L_2+1} \left(\frac{Z_1 Z_2}{\bar{Z}_1\bar{Z}_2}\right)^{{J}_{3;a}} \,, 
\end{aligned}
\end{align}
and analogue simplifications hold in presence of more excitations 
(see \cite{Olivucci:2023tnw} for a detailed discussion). This observation allows 
us to fix the integration contour with a prescription that makes sense of 
the decoupling effect in relation with the kinematical region, i.e. the 
signature of integration. First, the decoupling pole must be avoided by 
adding a small imaginary shift to the integration contour
\begin{equation}
\Gamma\left (-{\bf i}u+{\bf i}v+\frac{a-b}{2} \pm\varepsilon\right)\,.
\end{equation}
The sign of the shift is fixed according to a \emph{physical} prescription, as for propagators in QFTs. 
The correct choice is $+\varepsilon$, as it is argued in appendix D.2 of \cite{Basso:2018cvy}. Let's see why: schematically, the SoV integral has the form
\begin{equation}
\int du dv |Z_1|^{{\bf i} u} |Z_2|^{{\bf i} v} \Gamma\left ({ \bf i} v-{\bf i} u \pm \varepsilon\right) f(u,v) \sim \int du dv  \frac{e^{{\bf i} u \log |Z_1| }e^{{\bf i} v \log |Z_2|}}{{\bf i} v -{\bf i} u \pm \varepsilon} f(u,v)\,,
\end{equation}
where $f(u,v)$ is regular at $u=v$. With the change of variables $u_{\pm} = \frac{1}{2}(v\pm u)$ we find
\begin{equation}
\int du_+ du_- \frac{e^{{\bf i} u_+ \log |Z_1||Z_2| }e^{{\bf i}  u_- \log |Z_2|/|Z_1|}}{{\bf i}  u_- \pm \varepsilon} F(u_+,u_-) \propto\!\! \int du_+ e^{{\bf i}  u_+ \log |Z_1||Z_2| } F(u_+,0)\!\! \int du_- \frac{e^{{\bf i}  u_- \log |Z_2|/|Z_1|}}{{\bf i}  u_- \pm \varepsilon}\,.
\end{equation}
The integration in $du_-$ translates the prescription $\pm \varepsilon$ into to the condition $|Z_1|>|Z_2|$ or $|Z_1|<|Z_2|$ in order to pick (or not) the residue at the decoupling pole.
\begin{figure*}[t]
\centering
\includegraphics[scale=0.65]{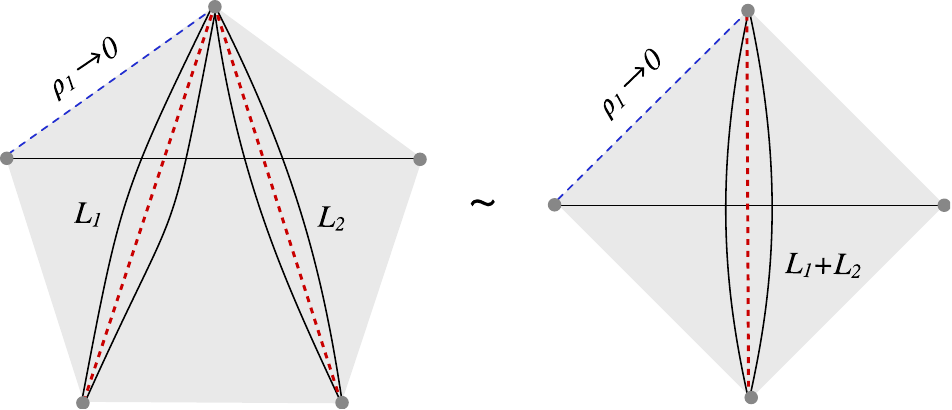}
\caption{A 5pt split-ladder with $L_1$, $L_2$ propagators along the cut in the kinematical limit of $\rho_1 = |Z_1|\ll1$ behaves as a ladder of length $L_1+L_2$. At the level of SoV integration, this effect is accounted for by the decoupling pole.}
\label{deca_kin}
\end{figure*}
Consider a $5$pt split-ladder and focus on its leading UV behaviour when $|Z_1|\ll1$ 
(or, similarly, when $|Z_2|\gg 1$). It is clear that such an integral with $L_1$, $L_2$ 
propagators along the cuts diverges as a logarithm $\sim \log^{L_1+L_2} |Z_1|$ 
plus subleading terms, ie. it shares the leading behaviour with a ladder with $L_1+L_2$ 
propagators, see FIG.\ref{deca_kin}. But this is exactly what \eqref{the_decoupling} 
accounts for! This observation fixes the correct prescription to be such that 
whenever $|Z_1|<|Z_2|$ the decoupling pole must fall inside 
the closure of the the integration contours, that is $+ \varepsilon$.

\section{Light-cone Stampedes fundamentals}

\begin{figure}[t]
\includegraphics[scale=0.9]{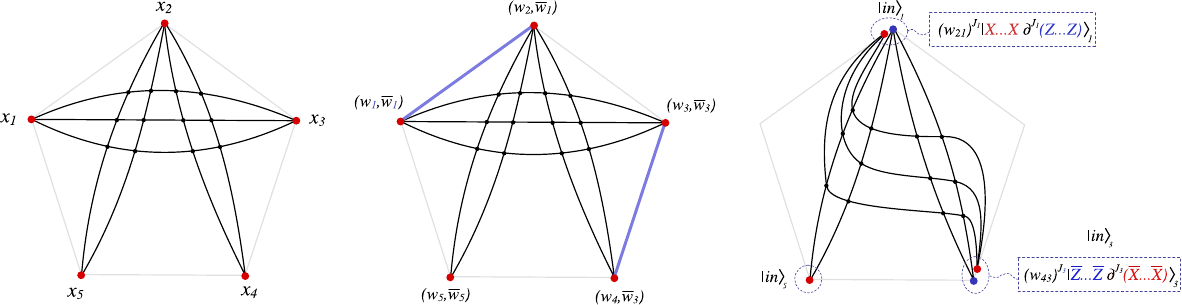}
\caption{The $5$-pt Fishnet $\mathcal{F}_{L_1,L_2,M}$ with $L_1=2=L_2$ and $M=3$, and its light-cone limit where blue edges lie on a light-cone ray in direction $w$, i.e. they are separated by zero distance in the light-cone direction $\bar w$. On the right, the illustration of the Taylor expansion that generates the stampedes initial state of equation \eqref{init_app_1} by replacing the $w$-direction displacements with fields at coinciding points and the insertion of light-cone derivatives.}
\label{5pt_stamp}
\end{figure}

The ideas behind the stampedes combinatorics, and the corresponding formulae that 
we derived in this letter, provide a \emph{new} way to explore the multi-point fishnet integrals 
for what concerns their leading behaviour in various UV limits. Let us  consider the 
limit where a distance along the $n$gon contour, say $x_{i,i+1}^2$, approaches the light-cone. 
In this regime a multi-point fishnet with $\ell$ integration points (\emph{loops}) develops 
a logarithmic divergence whose leading term takes the form
\begin{equation}
\label{OPE_logs}
\sim \log^n x^2_{i,i+1}\,,\,\,\, n \leq \ell \,.
\end{equation}
An instructive example, to be analyzed in the following, is that of a 5-pt fishnet diagram 
$\mathcal{F}_{L_1,L_{2},M}$ on the plane, like the one in Fig.\ref{5pt_stamp}, for which $\ell =L_1 M+L_2 M$. 
Introducing the conformal invariants, defined in agreement with \eqref{crossratios}, 
\begin{equation}
\label{5pt_ratios}
Z_1 \bar Z_1 = \frac{x_{12}^2 x_{45}^2}{x_{15}^2 x_{24}^2}\,\,\, ; \,\,\, (1-Z_1) (1-\bar Z_1) = \frac{x_{14}^2 x_{25}^2}{x_{15}^2 x_{24}^2} \,\,\, ; \,\,\, Z_2 \bar Z_2 = \frac{x_{25}^2 x_{34}^2}{x_{23}^2 x_{45}^2}\,\,\, ; \,\,\, (1-Z_2) (1-\bar Z_2) = \frac{x_{24}^2 x_{35}^2}{x_{23}^2 x_{45}^2}\,,
\end{equation}
we will study the diagram in planar kinematics, i.e. parametrizing the point $x_j$ by a 
pair of light-cone coordinates $(w_j,\bar w_j)$ such that $x_j^2 = w_j \bar w_j$.
The maximal degree of the log-divergence for a given light-cone limit can be read 
straightforwardly from the topology of the Feynman diagram. Indeed, 
the number $n$ in \eqref{OPE_logs} is equal to the number of those vertices 
that lie at the intersection between propagators emitted at points $x_i$ and $x_{i+1}$ 
-- see for reference FIG.\ref{5pt_stamp} --  and can be regarded as being related 
to the quantum corrections to the anomalous dimension of the operators flowing 
in the OPE channel $x_{i,i+1}^2 \sim 0$. Following such a prescription, 
\begin{align}
\begin{aligned}
\label{table_OPE}
\begin{tabular}{ |p{2cm}||p{3cm}|}
 \hline
$i$ & leading behaviour \\
 \hline
$1$ & $(\log x_{12}^2)^{(L_1+L_2)M} $  \\
$2$ &  $(\log x_{23}^2)^{(L_1+L_2)M}$   \\
$3$ & $(\log x_{34}^2)^{L_2 M}$   \\
$4$  & $1$  \\
$5$  & $(\log x_{15}^2)^{L_1 M}$   \\
 \hline
\end{tabular}
\end{aligned}
\end{align}
is the result we find for $\mathcal{F}_{L_1,L_{2},M}$.\\

So far we have considered a local behaviour of the integral around the points 
$x_i\,,x_{i+i}$ that are the locations of a product of $N$ and $N'$ fields of 
Fishnet CFT $\phi_j, \phi'_j \in \{\mathcal{X}, \mathcal{Z},\bar{\mathcal{X}}, \bar{\mathcal{Z}} \}$, say
\begin{equation}
\label{a}
O(x_i)=\phi_1(x_i)\cdots \phi_{N}(x_i)\,,\, O'(x_{i+1})=\phi'_1(x_{i+1})\cdots \phi'_{N'}(x_{i+1})\,.
\end{equation}
Assuming that the light-cone is approached on the plane as $\bar w_{j+1}\to \bar w_{j}$, 
we shall write the product of operators as a Taylor-expansion in the null direction $w_{j+1,j}=w_{j+1}- w_{j}$, namely
\begin{equation}
\label{OPE_reloaded}
O(w_j,\bar w_j)O'(w_{j+1},\bar w_{j})= O(w_j,\bar w_j)e^{w_{j+1,j}\cdot \partial_{j}} O'(w_j,\bar w_{j})  \,= \,\sum_{k=0}^{\infty}  \frac{w_{j+1,j}^k}{k!} O(w_j,\bar w_j) \partial^k_j O'(w_j,\bar w_j)\,,
\end{equation}
This Taylor expansion in \eqref{OPE_reloaded} differs from the 
standard OPE because it is made of operators that are not eigenstates 
of the dilations, and it has the advantage that the 
three-point tensor structure is tree-level. This type of expansion is convenient 
when focussing on the leading-log divergencies of a multi-point correlator, 
which are insensitive to the quantum correction of the three-point structure-constants. 
A short-distance limit $x_{i,i+1}^2 \to 0$ in a correlator probes the quantum corrections 
generated by the action of the dilations on the Fock space generated 
in $x_i$ by the operators \eqref{OPE_reloaded} on a vacuum,
\begin{equation}
\label{OPE_states}
(\phi_1\cdots \phi_{N})(w_i,\bar w_i) \partial_i^k (\phi'_1\cdots \phi'_{N'})(w_i,\bar w_i) |\text{vacuum}\rangle \equiv \sum_{h_1+\dots h_{N'}=k} |\phi_1, \dots,  \phi_{N}, \hat \partial^{h_1} \phi'_1 ,\dots, \hat \partial^{h_{N'}} \phi'_{N'} \rangle_i\,,
\end{equation} 
where we made us of the compact notation $\hat \partial_i^k \equiv \tfrac{1}{k!} \partial^k/\partial w_i^k$.
In particular, the leading term at loop-order $\ell$ results from the $\ell$-fold action of the one-loop dilation operator on the states \eqref{OPE_states}. In the Fock space formalism such action is a nearest-neighbour interaction $\mathbb{H}_i$ between the sites of a chain with open boundaries
\begin{equation}
\label{hopping}
\mathbb{H}_i \, |\hat \partial^{h_1} \phi_1, \dots , \hat \partial^{h_m} \phi_m \rangle_i = \sum_{k=1}^{m-1} \hat{h}_{k,k+1} \,|\hat  \partial^{h_1} \phi_1, \dots ,  \hat \partial^{h_m} \phi_m \rangle_i  \,.
\end{equation}
In the Fishnet theory the only non-zero matrix elements of the local \emph{hopping} operator $\hat{h}_{1,2}$ are given, for any integers $n,m$, by the following map
\begin{equation}
\label{nn_local}
\hat h_{12}\, | \hat \partial^{n} \mathcal X , \hat \partial^{m} \mathcal Z\rangle_i = \!\frac{1}{n+m+1}\! \sum_{j=0}^{n+m} \,|\hat  \partial^{n+m-j}  \mathcal Z , \hat \partial^{j} \mathcal X\rangle_i\,,
\end{equation}
and by the elements obtained upon cyclic permutations $\mathcal X\to \mathcal  Z \to \bar{\mathcal X} \to \bar{\mathcal Z}$. In particular, due to the \emph{chirality} of the interaction, the opposite ordering of fields is always annihilated:
\begin{equation}
\hat h_{12}\, |\hat  \partial^{n} \mathcal Z , \hat \partial^{m} \mathcal X \rangle_i =0\,.
\end{equation}
On the top of that, the one-loop dilations $\mathbb{H}_i$ redistributes light-cone 
derivatives $\hat \partial$ between neighbouring fields, ie. $sl(2)$ excitations 
in the $\infty$-dimensional irrep of spin $s=-1$. The specific form of the action 
\eqref{nn_local} can be read out of the OPE expansion of a conformal Box integral.\\

The description of light-cone leading-logarithms given so far explains the orders 
of divergence in the table \eqref{table_OPE} as the maximal number of actions 
of $\mathbb{H}_i$ on the Fock-space states produced by the expansion \eqref{OPE_reloaded} 
around the point $x_i$. Take for instance $i=1$; the OPE states are of the type
\begin{equation}
 | \underbrace{\mathcal X , \dots ,\mathcal X}_{M} , \underbrace{\hat \partial^{h_1} \mathcal Z ,\dots ,\hat  \partial^{h_{L_1}} \mathcal Z}_{L_1}, \underbrace{\mathcal Z, \dots ,\mathcal  Z}_{L_2}  \rangle\,,
 \end{equation}
 which after $(L_1 + L_2)M$ actions takes the form
 \begin{equation}
 |\underbrace{\hat \partial^{h'_1} \mathcal Z ,\dots , \hat \partial^{h'_{L_1+L_2}} \mathcal Z}_{L_1+L_2}, \underbrace{\mathcal X, \dots ,\mathcal  X}_{M}  \rangle\,,
 \end{equation}
and are annihilated by any further application of $\mathbb{H}_i$ as a consequence of chirality of the Fishnet CFTs \cite{Gurdogan:2015csr}.\\ 

The idea at the basis of the \emph{stampede} method here is that, for a given fishnet diagram, 
there is a choice of simultaneous limits $x_{i,i+1}^2 \to 0$ that develop a maximal log-divergence, 
i.e. when the total power of logs matches the loop order.
%
Let us elaborate further on this point and let's take a set of indices 
$j \in P \subset \{1,\dots ,n\}$ for an multi-pt fishnet diagram with $\ell$ vertices, and for every $j$ introduce a set of consecutive indices
\begin{equation}
S_j =\{j,j+1,\dots, j+m_j\}\,,
\end{equation}
constrained to be disjoint, $S_j \cup S_i = \varnothing$.
When all the points $x_{i}=(w_i,\bar w_i),\, i\in S_j$ lie on a light-ray --  say $\bar w_{j,i}=0$ -- the diagram develops UV divergences in the form of a logarithmic behaviour which is stratified into powers of $\lambda_j \sim -\log \bar w_{k,j} \sim -\log \bar w_{h,j}$ for $h,k\in S_j$. As explained before, the leading term is expected to be $\lambda^{n_j}$ where $n_j$ is the number of vertices in the diagram that mark the intersection of two lines emitted by points $x_i,x_k$ both on the same light-ray, namely $i,k \in S_j$. When the light-cone limit is approached simultaneously for all $j \in P$ the diagram develops a maximal log-divergence iff 
\begin{equation}
\label{maxlog_condition}
n_1 +n_2 +\dots +n_p \geq \ell\,.
\end{equation}
It shall be clear that the sign $>$ in \eqref{maxlog_condition} implies 
that at least one vertex is at the intersection of two lines that have both 
pairs of extremal points aligned on a light-ray, see also FIG.\ref{ABC_graphs} (B). 
\begin{figure}[b]
\includegraphics[scale=0.95]{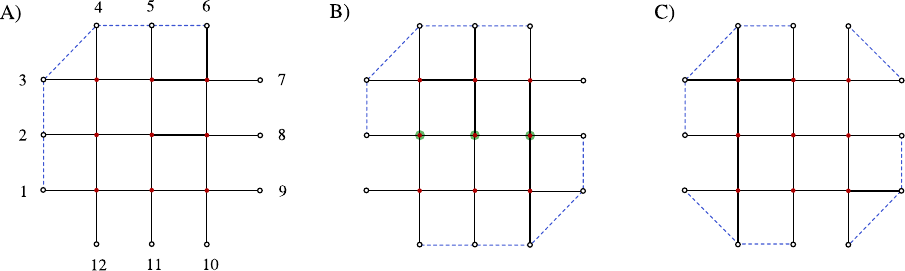}
\caption{A fishnet Feynman integral with $9$ loops and $n=12$ external points, in principle all different. Blue dashed lines group points that are aligned on a light cone direction, say setting equal $\bar w$ on the plane. We depict three kinematical limits where the stampede method is applicable. (A): $P=\{1\}$ and $S_1=\{1,\dots, 6\}$. (B): $P=\{1,8\}$ and $S_1=\{2,\dots, 6\}$, $S_2=\{8,\dots, 12\}$. (C): $P=\{2,6,8,11\}$ and $S_1=\{2,\dots, 5\}$, $S_2=\{6,7\}$, $S_3=\{8,10\}$, $S_4=\{11,12,1\}$. In the case (B) we enlighten with a circle such vertices that lie at the crossing of pairs of lines emitted by points in both $S_1$ and by $S_2$.}
\label{ABC_graphs}
\end{figure}
Hence, in general the maxlog is given by a stratification into monomials of homogeneous degree
\begin{equation}
\prod_{j \in P} \lambda_j^{k_j}\qquad;\qquad\sum_{j \in P} k_j = \ell\,.
\end{equation}
Notice that since we are interested in the maxlogs of a fishnet diagram, the minimal number
$|P|$ of subsets $S_j$ one should consider is one, as in the example of 
FIG.\ref{ABC_graphs} (B), while, due to the shape of a fishnet diagram, 
the maximal number is $|P|=4$ (FIG.\ref{ABC_graphs}, (C)).\\

Define $S\equiv \cup_{j} S_j \backslash\{x_j\}$.
Crucially, the indices $i \in S$ 
must be chosen such that \emph{all} the integrated points of the fishnet 
graph lie at the intersection of two lines emitted by any of the pair
of points $x_{i},\, i \in S_j$ for some $j$. Under these circumstances the 
quantum corrections at loop order $\ell$ are realized by the $\ell$-fold action 
of the one-loop dilations and the correlator factorizes into divergent logs times 
an analytic function of the displacements $\{w_{i,j},\, i \in S_j\}$ that is relatively simple to compute as a Taylor series:
\begin{equation}\label{idea_logs}
\mathcal F =  \sum_{|\underline k|=\ell}\left[\left( \prod_{i \in P} \lambda_i^{k_i}\
\right) \times \sum_{\underline{J}\geq 0} \mathbf{f}_{\underline{k};\underline{j}}(\eta,\bar \eta)\prod_{j \in P}\prod_{s \in S_j}w_{s,j}^{J_s} + \dots \right]+\dots\,,
\end{equation}
where $k_i$ denotes the power of repeated actions of $\mathbb{H}_{i}$ on the 
state \eqref{OPE_states} in $x_{i \in P}$, $(\eta,\bar \eta)$ stands for the dependence 
over all points $x_k=(w_k,\bar w_k)$ for $k\not \in S$, and $\dots$ denote any sub-leading 
term in the light-cone regime. The summand labelled by $k_1,k_2,\dots, k_{|P|}$ 
in \eqref{idea_logs} is associated with a certain \emph{stampede process} which we 
can generate systematically according to the procedure we are going to illustrate. 
First, one defines the initial state of the stampede. For an $n$-point fishnet diagram the 
initial state is the tensor product of $n-|S|$ initial vectors. Each vector sits at one of 
the space-time points left intact by the light-cone expansions \eqref{OPE_reloaded}, 
namely $x_j$ with $i \in P$, or in one of the points not involved in any light-cone 
condition $x_i$, $i\not \in \cup_j S_j$. We use the following notation:
\begin{equation}
\label{init_def}
| \text{in}(\underline{J})\rangle =  \bigotimes_{i\in \{1,\dots, n\}/S} | \text{in}(\underline{J}_i)\rangle_i \,,
\end{equation}
and for $i \in P$ this means
\begin{equation}
| \text{in}(\underline{J_i})\rangle_i  = O_{i} \hat{\partial}^{J_1} O_{i+1}(w_i,\bar w_{i}) \cdots  \hat{\partial}^{J_{m_i-1}} O_{i+m_i} (w_i,\bar w_{i}) |\text{vacuum}\rangle\,,
\end{equation}
while for all the indices $i\not \in \cup_j S_j $ the states has $J_i =0$ and is simply given by
\begin{equation}
| \text{in}\rangle_i = O_i(w_i,\bar w_i)| \text{vacuum}\rangle\,.
\end{equation}
Next, a $n$-pt fishnet diagram is defined by $N\geq n$ external fields, 
located in positions $\{y_1,\dots ,y_N\}$, where $y_k \in \{x_1,\dots, x_n\}$. 
Thus, the states of the stampedes, for instance the one in \eqref{init_def}, are a tensor 
product of $N$ single-field vectors, eventually located at coinciding points. We 
define the functional $\mathbf{C}$ acting on such tensor products as the free-theory 
contraction of fields $\partial^{J_i}  \phi_i(y_i)$ and $\partial^{J'_i}  \phi_i'(y'_i)$ that 
stand at endpoints of a given line in the fishnet lattice, namely
\begin{equation}
\label{C_contra}
\mathbf{C} \cdot \bigotimes_{i=1}^{N} |\hat \partial^{J_i} \phi_i\rangle_{y_i}  \!= \!\prod_{i=1}^{N/2}\langle \hat \partial^{J'_{i}} \phi'_{i}(y'_{i}) \hat  \partial^{J_i} \phi_i(y_i) \rangle_{\text{free}}\,.
\end{equation}
%
%

The generating function of leading-logarithms $\mathcal{G}_{\underline{J}}(\underline{\lambda})$ 
is obtained as the result of evolving the initial state into a final state at times $\lambda_k$ 
followed by the contraction of the final state
\begin{equation}
| \text{in}(\underline J)\rangle \, \mapsto \, \mathbf{C}\cdot | \text{out}(\underline \lambda; \underline J)\rangle = \mathbf{C} \cdot e^{\sum_{i\in S} \lambda_i \mathbb{H}_i}   | \text{in}\left(\underline J \right)\rangle \,.
\end{equation}
A given \emph{stampede function} is extracted as the corresponding
monomial in the generating function with respect to the times $\lambda_k$, i.e.
\begin{equation}
\label{stampede_func_gen}
\mathbf{f}_{\underline{k}; \underline{J}}(\eta,\bar \eta) =\left(\prod_{i\in S} \frac{d^{k_i}}{d\lambda_i^{k_i}}\right) \mathbf{C}\cdot | \text{out}(\underline{\lambda}; \underline{J})\rangle \left. \right|_{\lambda_i=0}\,.
\end{equation}
\begin{figure}[t]
\includegraphics[scale=0.95]{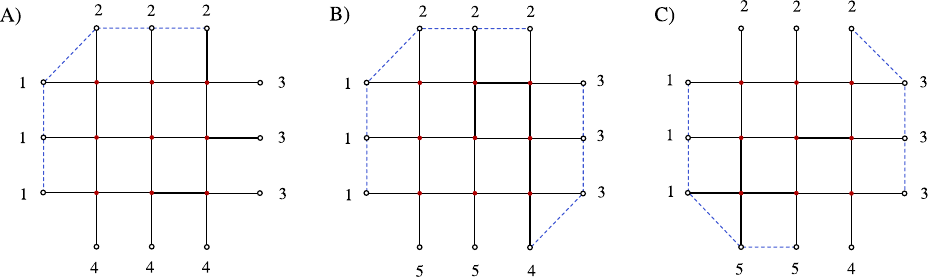}
\caption{Examples of light-cone limits that generate maxlogs in a 4pt fishnet integral (A), and 5pt fishnet integral (B),(C). The stampede (A) results in the vertex model partition function discussed in the body of the letter. The cases (B) and (C) generate respectively the maxlogs in formulae \eqref{5pt_stamp_I} and \eqref{5pt_stamp_II}, when $M_1=M_2=3$ and $L_1=2$, $L_2=1$.}
\label{ABC_graphs}
\end{figure}

\begin{figure}[t]
\includegraphics[scale=0.9]{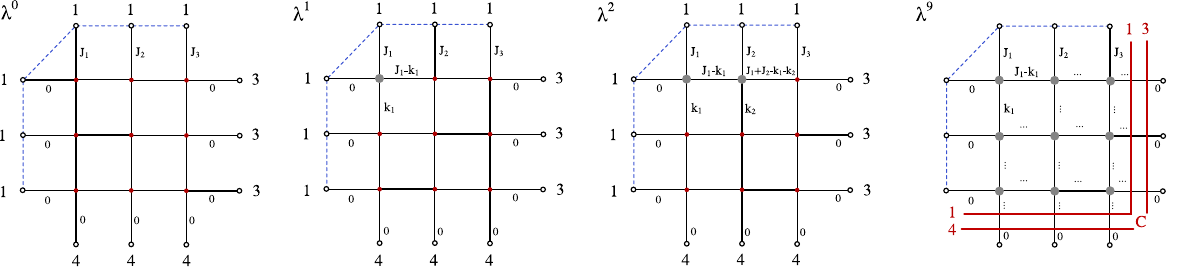}
\caption{The generation of a vertex model partition function that describes the maxlog in the limit $\bar w_2\to \bar w_1$ of the $9$ loop $4$pt Basso-Dixon fishnet integral, represented in Fig.\ref{ABC_graphs} (A).  We illustrate the stampede process from left to right at times $0$, when the state $| \text{in}(J)\rangle_1$ is generated, with $J_1+J_2+J_3=J$. Then, after one and two actions of $\mathbb{H}_1$ the derivatives start to ``flow" in the lattice from up-left to down-right. Finally, at the final step one takes the free-theory contraction $\mathbf C$ between the states at points $(w_1,\bar w_1), (w_3,\bar w_3)$, and $(w_4,\bar w_4)$.}
\label{4p_deriv}
\end{figure}

In the body of the letter we mentioned explicitly the $4$pt fishnet maxlog as derived 
from \eqref{stampede_func_gen}.   Let us explain this derivation in more details, referring 
to Fig.\ref{4p_deriv}, that is the fishnet integral with $L_1=M_1=3$. In the first picture, 
at ``time" zero the initial state is generated by insertion of derivatives, i.e. light-cone Taylor expansion $w_2\sim w_1$
\begin{equation}
| \text{in}(J)\rangle_1 = \sum_{J_1+J_2+J_3=J}| \mathcal{X ,X ,X} , \hat \partial^{J_1} \mathcal{Z},   \hat \partial^{J_2} \mathcal{Z},  \hat \partial^{J_3} \mathcal{Z}  \rangle_1\,.
\end{equation}
Then, the one-loop action of $\mathbb{H}_1$ inserts the first quantum correction to 
the initial state, ie.~it ``switches on" the up-left vertex, hence marked by a grey blob,
\begin{equation}
\mathbb{H}_1| \mathcal{X ,X, X} , \hat \partial^{J_1} \mathcal{Z} , \hat \partial^{J_2} \mathcal{Z} , \hat \partial^{J_3} \mathcal{Z}  \rangle_1 = \frac{1}{J_1+1} \sum_{k_1=0}^{J_1}  |  \mathcal{X,X} ,\hat \partial^{k_1}  \mathcal{Z} , \hat \partial^{J_1-k_1} \mathcal{X} , \hat \partial^{J_2} \mathcal{Z} , \hat \partial^{J_3} \mathcal{Z}  \rangle_1 \,.
\end{equation}
At time $\lambda^2$, the action of $\mathbb{H}_1$ adds the correction corresponding 
to the second vertex in the first row, as depicted in FIG.\eqref{4p_deriv}, or, alternatively, to the second vertex in the first column. Indeed, in formulae,
\begin{align*}
\begin{aligned}
&\mathbb{H}_1 |\mathcal{X,X} ,\hat \partial^{k_1}  \mathcal{Z} , \hat \partial^{J_1-k_1} \mathcal{X} , \hat \partial^{J_2} \! \mathcal{Z} , \hat \partial^{J_3} \! \mathcal{Z}  \rangle_1  = \\
&=(\mathbbm{h}_{23}+\mathbbm h_{45}) |\mathcal{X,X} ,\hat \partial^{k_1} \! \mathcal{Z} , \hat \partial^{J_1-k_1} \! \mathcal{X}, \hat \partial^{J_2} \mathcal{Z} , \hat \partial^{J_3} \mathcal{Z}  \rangle_1  =\\
&=\frac{1}{k_1+1}\sum_{h_1=0}^{k_1} |\mathcal{X} , \hat \partial^{h_1}  \! \mathcal{Z} ,  \hat \partial^{k_1-h_1}\!  \mathcal{X}  , \hat \partial^{J_1-k_1}\! \mathcal{X}  , \hat \partial^{J_2} \!\mathcal{Z} ,  \hat \partial^{J_3}\! \mathcal{Z}  \rangle_1 +\\ &+\frac{1}{J_1+J_2-k_1+1} \sum_{k_2=0}^{J_1+J_2-k_1} |\mathcal{X,X} , \hat \partial^{k_1} \! \mathcal{Z} , \hat \partial^{k_2} \!\mathcal{Z}  , \hat \partial^{J_1+J_2-k_1-k_2}\! \mathcal{X}  , \hat \partial^{J_3}\! \mathcal{Z}  \rangle_1  \,.
\end{aligned}
\end{align*}
 At the final time in FIG.\eqref{4p_deriv}, ie.~at order $\lambda^9$, one is left with three states. The evolution at time $9$ of the initial state, carrying certain numbers of derivatives symbolized with ``\ldots" on its sites, all localized in $(w_1,\bar w_1)$, and other two states. 
 The other two states do not contain light-cone derivatives, and sit at points $(w_3,\bar w_3)$ 
 and $(w_4,\bar w_4)$. At this point one performs the free-theory contractions $\mathbf C$ site-by-site. 
 It is simple to check that, in the convenient conformal frame $(w_1 , \bar w_1) =(0,0)$, 
 $(w_2 , \bar w_2) =(Z,\bar Z)$, $w_3=\bar w_3 =\infty$ and $(w_4 , \bar w_4) =(1,\bar 1)$, 
 the operation of contraction simplifies remarkably and the partition function of FIG.\ref{pf_BD} 
 in the main text is straightforwardly recovered.\\

For clarity we will give two more very concrete examples for $5$pt fishnet diagram with $\ell =(L_1+L_2)M$ loops, 
parametrized on the plane by points $x_i=(w_i,\bar w_i)$ and by the conformal invariants \eqref{5pt_ratios}. 
First, we take the limit $x_{12}^2, x_{34}^2 \to 0$ in $\bar w$ direction
\begin{equation}
\bar w_{2} \to \bar w_1\, ,\,\, \bar w_{4} \to \bar w_3 \,,
\end{equation} 
which, in the terminology introduced at the beginning of the section, corresponds to $P=\{1,3\}$ and $S_1 = \{1,2\},\, S_3 = \{3,4\}$.
The divergent logarithms can reach maximal powers $\lambda_1^{(L_1+L_2)M}$ and $\lambda_1^{M L_2}$ respectively, and generally there are mixed terms such that the total powers equals $M(L_1+L_2)$, hence the leading behaviour of this fishnet diagram reads
\begin{equation}
\label{5pt_stamp_I}
\sum_{k=0}^{M L_2}\lambda_1^{M L_1 +k} \lambda_3^{M L_2 -k} \times \sum_{J_1,J_3\geq 0} \mathbf{f}_{M L_1 +k,M L_2 -k;J_1,J_3}(\eta,\bar \eta) \, w_{21}^{J_1}  w_{43}^{J_3}\,,
\end{equation}
where $(\eta,\bar \eta) = \{(w_1,\bar w_1),\, (w_3,\bar w_3),\,(w_5,\bar w_5)\}$.
The initial states of the process $|\text{in}(J_1)\rangle_{1} \otimes |\text{in}(J_3)\rangle_3 \otimes |\text{in}\rangle_5$ are
\begin{align}
\begin{aligned}
\label{init_app_1}
&|\text{in}(J_1)\rangle_{1}  = | \underbrace{\mathcal X \cdots \mathcal X}_{M}\rangle_{1} \otimes \hat \partial^{J_1} |  \underbrace{\mathcal Z \cdots \mathcal Z}_{L_1+L_2}  \rangle_{1}\\ 
&|\text{in}(J_3)\rangle_{3}  = |  \underbrace{\bar{\mathcal X} \cdots \bar{\mathcal X}}_{M}  \rangle_{3} \otimes \hat \partial^{J_3} |  \underbrace{\bar{\mathcal Z} \cdots \bar{\mathcal Z}}_{L_2}  \rangle_{3}\\
&|\text{in}\rangle_{5}  =  |  \underbrace{\bar{\mathcal Z} \cdots \bar{\mathcal Z}}_{L_2}  \rangle_{5} \,.
\end{aligned}
\end{align}
The generating function of the leading-logs mixes the action on $x_1$ and $x_3$ giving rise to a few monomials in $\{\lambda_1,\lambda_3\}$,
\begin{equation}
\mathcal{G}_{J_1,J_3}(\lambda_1,\lambda_3)=\mathbf{C}\cdot \left(e^{\lambda_1 \mathbb{H}_1}  |\text{in}(J_1)\rangle_{1} \otimes e^{\lambda_3 \mathbb{H}_3} | \text{in}(J_3)\rangle_{3} \otimes | \text{in}\rangle_{5}   \right)\,.
\end{equation}
Following \eqref{stampede_func_gen} different stampede functions are associated to different processes.  For instance, $k=M L_2$ corresponds to the action of $\mathbb{H}$ exclusively on the state $|\text{in}\rangle_{1}$
\begin{equation}
|\text{out}(J_1)\rangle_{1}  =\mathbb{H}_1^{M(L_1+L_2)} |\text{in}(J_1)\rangle_{1} \,,\,\,\, |\text{out}\rangle_{3,5} =|\text{in}\rangle_{3,5}  \,,
\end{equation}
and the stampede function is
\begin{equation}
\mathbf{f}_{M(L_1+L_2),0;J1,J3}= \mathbf{C}\cdot \left( \mathbb{H}_1^{M(L_1+L_2)} |\text{in}(J_1)\rangle_{1} \otimes | \text{in}(J_3)\rangle_{3} \otimes | \text{in}\rangle_{5}   \right) \,.
\end{equation}
Another instance is the process $k=0$, where the action is distributed at among points $x_1$ and $x_3$,
\begin{equation}
|\text{out}(J_1)\rangle_{1}  =\mathbb{H}_1^{M L_1} |\text{in}(J_1)\rangle_{1} \,,\, |\text{out}(j_3)\rangle_{3}  =\mathbb{H}_3^{M L_2} |\text{in}(J_3)\rangle_{3} \,,\, |\text{out}\rangle_{5} =|\text{in}\rangle_{5}  \,,
\end{equation}
and the stampede function reads
\begin{equation}
\mathbf{f}_{M L_1,M L_2;J1,J3}= \mathbf{C}\cdot \left(\mathbb{H}_1^{M L_2}|\text{in}\rangle_{1} \otimes \mathbb{H}_3^{M L_2}  |\text{in} \rangle_3\otimes | \text{in}\rangle_5\right)\,.
\end{equation}
Another examle is when the same $5$-pts fishnet diagram of FIG.\ref{5pt_stamp}, is subject to a different light-cone limit 
\begin{equation}
\bar w_{1} \to \bar w_5 \,,\,\bar w_3 \to \bar w_4 \,,
\end{equation} 
that is $P=\{5,3\}$ and $S_3=\{3,4\}\,,S_5=\{5,1\}$. The divergent logarithms can be at most $\lambda_5^{M L_1}$ and $\lambda_3^{M L_2}$ respectively; since the total powers equals the loop order $\ell= M(L_1+L_2)$ the leading behaviour in this case is described by one monomial only
\begin{equation}
\label{5pt_stamp_II}
\lambda_5^{M L_1} \lambda_3^{M L_2} \times \sum_{J_3,J_5\geq 0} \mathbf{f}_{M L_2,M L_1 ;J_3,J_5}(\eta,\bar \eta) \, w_{15}^{J_5}  w_{43}^{J_3} \,.
\end{equation}
The initial states of the process is $|\text{in}(J_5)\rangle_{5} \otimes |\text{in}\rangle_2 \otimes |\text{in}(J_3)\rangle_3$, that is, more explicitly:
\begin{align}
\begin{aligned}
&|\text{in}\rangle_{5}  = | \underbrace{\bar{ \mathcal Z} \cdots \bar{\mathcal Z}}_{L_1}\, \rangle_{5} \otimes \hat{\partial}^{J_5} |  \underbrace{\mathcal X \cdots \mathcal X}_{M} \, \rangle_{5}\,,\,\,\,|\text{in}\rangle_{2}  =  |  \underbrace{\mathcal Z \cdots  \mathcal Z}_{L_1+L_2}  \rangle_{2}\,,\\
&|\text{in}\rangle_{3}  = |  \underbrace{\bar{ \mathcal X} \cdots \bar{\mathcal X}}_{M}  \rangle_{3} \otimes \hat{\partial}^{J_3} |  \underbrace{\bar{\mathcal Z} \cdots \bar{\mathcal Z}}_{L_2}  \rangle_{3}\,,
\end{aligned}
\end{align}
and the only non-trivial stampede function is
\begin{equation}
\mathbf{f}_{M L_2,M L_1;J_3,J_5}= \mathbf{C}\cdot \left( | \text{in}\rangle_2 \otimes \mathbb{H}_3^{M L_2}|\text{in}(J_3)\rangle_{3} \otimes \mathbb{H}_5^{M L_1}  |\text{in}(J_5) \rangle_5\right)\,.
\end{equation}


\section{Four-point fishnets reloaded: a Cauchy identity bootstrap}

The SoV representation of the 4-pt fishnets, in our notation reads,
\begin{align}
{\cal F}_{L,M}&=\mu_{\bf a}({\bf u}) \, E_{\bf a}({\bf u})^{L+M}\left(Z\bar{Z}\right)^{\sum_{k=1}^{M} \left({\bf i} u_{k} -\frac{1}{2}\right)} \,
 \text{Tr}_{\mathbf{a}}\left[\otimes_{k=1}^{M} \left(\frac{Z}{\bar{Z}}\right)^{\!\!{J}_{ z;a_{k} }}\right] =\label{BD_inte}\\[.2cm]
 &=  \frac{1}{(2\pi {\bf i})^M M!}\frac{ \rho^{\sum_{k=1}^{M}\! s_k} }{ \prod_{j=1}^{M} (-s_j)^{L+M} } 
 \frac{  {\rm VdM}({\bf s}){\rm VdM}({\bf s+a}) \,\prod_{j,k}\!\big(s_k-(s_j+a_j)\big) }{ \prod_{k=1}^{M} (a_k+s_k)^{L+M} }\times h_{\bf a}(Z,\bar{Z}) \notag
\end{align}
where in the second line we substituted ${\bf i}\, u_j= \tfrac{a_j}{2}+s_j$ 
and introduced the notation ${\rm VdM}({\bf s})=\prod_{k<j} (s_k-s_i)$. We also defined,
\begin{equation}
h_{\bf a}(Z,\bar{Z})=\rho^{\sum_{k=1}^{M}\!\frac{a_k-1}{2} }\times 
\text{Tr}_{\mathbf{a}}\left[\otimes_{k=1}^{M} \left(\frac{Z}{\bar{Z}}\right)^{\!\!{J}_{ z;a_{k} }}\right]
= \prod_{j=1}^M h_{a_j}(Z,\bar{Z})\qquad;\qquad h_{a\ge 1}=\frac{Z^{a}-\bar{Z}^{a}}{Z-\bar{Z}}
\end{equation}
where $h$ is the standard homogeneous symmetric polynomial. 
Note that what was the term $\prod_i a_i$ in $\mu_{\bf a}$ \eqref{defmu} now has gone into $\prod_{j,k}\!\big(s_k-(s_j+a_j)\big)$
since the index run over the full square $1\leq j,k\leq M$.\\

In the following we will compute the maximal $\log$ discontinuity of ${\cal F}_{L,M}$
and show that it naturally takes the form of a determinant of a matrix whose entries 
are maximal logs of ladders. Steinman relations will imply that ${\cal F}_{L,M}$ is a 
determinant of ladder, as in the original discussion by Basso and Dixon \cite{Basso:2017jwq}.\\

There are only higher degree poles at $s_j=0$ to be considered. These 
can be turned into derivatives $\partial_{s_j}$, by using {\tt ibp} through the identity
\begin{equation}
\frac{1}{(-s)^{N+1}}=-\frac{1}{N!}\partial_s^{N}\left(\frac{1}{s}\right)\qquad;\qquad N\equiv L+M-1,
\end{equation} 
The degree of these poles is too high 
to get the allowed maximal power of $\log(\rho)$ by acting with all derivatives on $\rho^{\sum_{k=1}^{M}\! s_k}$.
The intuition is that a number of derivatives will have to be used to neutralise
the ${\rm VdM}({\bf s})$, which otherwise evaluates to zero at $s_j=0$.
To proceed more systematically, we will use the (dual) Cauchy identity for 
symmetric Schur polynomials $P_{\underline{\lambda}}$, see e.g.~\cite{Macdonald_book}, namely
\begin{align}\label{dual_C}
\sum_{\underline{\lambda}\subseteq n^m } P_{\underline{\lambda}}({\bf x}) P_{\underline{\lambda}^c}({\bf y})=
 \prod_{k=1}^{ m}\prod_{j=1}^{n} ( x_k-y_j)\qquad;\qquad 
 (\underline{\lambda}^c)_j= m-(\underline{\lambda}')_{n+1-j}.
\end{align}
We use \eqref{dual_C} to rewrite the square $\prod_{j,k} (s_k-(s_j+a_j))$ in terms of $x=s$ and $y=s+a$. Then, we obtain
\begin{align}
{\cal F}_{L,M}\ \underbracket{\ =\ }_{\textnormal{{\tt ibp}}}\ \frac{ (-)^{ML}}{ \prod_j N! }h_{\bf a}(Z,\bar{Z})
\times\!\!\!\sum_{\underline{\lambda}\subseteq M^M } {\cal T}_{L,M,\underline{\lambda}} 
\qquad;\qquad  
{\cal T}_{L,M,\underline{\lambda}}=\prod_{j=1}^{M} \partial_{s_j}^{N}\bigg[ \tilde{P}_{\underline{\lambda}}({\bf s})
\times \rho^{\sum_{k=1}^{M}\! s_k} Q_{\underline{\lambda}}({\bf s}+{\bf a}) \bigg]
\end{align} 
where $\tilde P({\bf s})\equiv{\rm VdM}({\bf s})P({\bf s})$ is the anti-symmetric Schur 
polynomial, and $Q({\bf s})$ is defined to contain all the rest. The precise form of $Q({\bf s})$ is not immediately important.\\

Now, focus on a Young diagram in the sum, call it  $\underline{\kappa}$. The anti-symmetric Schur polynomial 
has a determinantal structure (which is stronger than saying 
$\tilde{P}_{\underline{\kappa}}({\bf s})={\rm VdM}({\bf s})\times$symmetric polynomial$({\bf s})$). 
More precisely
\begin{equation}
\tilde{P}_{\underline{\kappa}}({\bf s})=\det\left[
\begin{array}{rcl}
s_1^{\kappa_1+M-1}  & \ldots & s_{1}^{\kappa_M+M-1} \\  
\vdots & & \vdots  \\ 
s_1^{\kappa_M} & \ldots & s_M^{\kappa_M}\end{array}\right]\,.
\end{equation}
Therefore, in order to have a non zero result when we act with derivatives 
and evaluate at $s_j=0$ there is only one possibility: The derivatives $\partial_{s_j}$ 
have to be distributed according to the partition $p_{j}=\kappa_j+(j-1)$ or a permutation of it. We find
\begin{align}
{\cal T}_{ L,M,\underline{\kappa} }=
 \prod_{j=1}^{M} \text{Bin}\Big[\,^{N}_{\,p_j}\Big]  p_j!\times
\sum_{\pi } \text{sgn}(\pi(p)) \prod_{j=1}^{M} \partial_{s_j}^{N-p_{ \pi(j) } } 
\bigg[ \rho^{\sum_{j=1}^{M}\! s_j}  Q_{ \underline{\kappa} } ({\bf s+ a})\bigg]_{ {\bf s}=0}
\end{align}
In this formula $p_j!$ is the result from the derivatives, and the binomial is 
counting the multiplicity of picking $p_j$ derivatives out of $N$.
At this point, we are left with $N-p_{ \pi(j) }$ derivatives for each $s_j$ 
to spend on the term in $\big[\ldots\big]$. Say we spend $q_j$ out of $N-p_{ \pi(j)}$ 
on $Q_{\underline{\kappa}}$, then the rest, ie.~$N-p_{\pi(j)}-q_j$, will act on $\rho$ 
generating a $\log(\rho)$ at $s_j=0$.  Thus
\begin{align}
{\cal T}_{ L,M,\underline{\kappa} }=\prod_{j=1}^{M} \text{Bin}\Big[\,^{N}_{\,p_j}\Big]  p_j!\times
\sum_{q_j} (\log\rho)^{MN-|p|-|q|} \times \bigg[ 
\sum_{\pi } \text{sgn}(\pi(p)) \prod_{j=1}^{M} \text{Bin}\Big[\,^{N-p_{\pi(j)}}_{\ \ \ q_j}\Big] \bigg] 
\partial_{s_1}^{q_1}\ldots\partial_{s_{M}}^{q_M} Q_{\underline{\kappa}}({\bf s+ a})\Bigg|_{{\bf s}=0}
\end{align}
We recognize
\begin{equation}\label{important_det}
 \bigg[ \sum_{\pi } \text{sgn}(\pi(p_{\underline{\kappa} } )) 
 \prod_{j=1}^{M} \text{Bin}\Big[\,^{N-p_{\pi(j)}}_{\ \ \ q_j}\Big] \bigg] =\det\Big[ {\rm Bin}\Big[\,^{N-p_i}_{\ \ q_j}\Big]_{1,\leq i,j\leq M} \Big]
\end{equation}
This determinant vanishes when some of the $q_j$ are equal, and it is equal 
to $1$ when $p=q=(M-1,\ldots,0)$ or permutations of that. For the maxlog we should consider the 
minimum $|p|$ and $|q|$. Note now that
we can only increase $p$ or $q$ starting from $(M-1,\ldots,0)$, and never decrease. 
If we decrease we fall into the case of $q_i=q_j$ for some $i\neq j$ and the determinant 
vanishes. This means that 
\begin{equation}
{\rm max}_{p,q}(MN-|p|-|q|)= M N-2\frac{M(M-1)}{2}= M (1 + N - M)= M L
\end{equation}
which is in fact half of the transcendental weight from the BD formula.

We have showed that $(\log\rho)^{ML}$ is the maximal log of the BD integral. Its coefficient function comes from 
 taking $\underline{\kappa}=\underline{0}$ 
in the Cauchy identity, ie.~$p=(M-1,\ldots)$, and  summing over $q_j$,
which again are permutations of $q^*=(0,\ldots,M-1)$. Thus,
\begin{equation}
{\cal I}_{L,M}\Big|_{(\log\rho)^{ML}}= (-)^{ML}h_{\bf a}(Z,\bar{Z})\times
\frac{ \prod_{i} \text{Bin}\Big[\!\,^{\ \ N}_{M-i}\,\Big]\times  (M-i)! }{\prod_i N! }
\sum_{\pi} \text{sgn}(\pi(q^* )) Q_{\underline{0}}^{(q^*_{\pi(1)},\ldots )}({\bf s+a})\Big|_{{\bf s}=0}
\end{equation}
At this point we need the explicit form of $Q_{\underline{0}}$ 
from \eqref{BD_inte}-\eqref{dual_C}, which reads
\begin{equation}
Q_{\underline{0}}({\bf s+a})=\frac{1}{M!}\,\frac{ {\rm VdM}({\bf s+a}) }{ \prod_i (a_i+s_i)^{L+M}} P_{[M,\ldots ,M]}({\bf s+a}) 
=\frac{1}{M!} \frac{ {\rm VdM}({\bf s+a}) }{ \prod_i (a_i+s_i)^L} 
\end{equation}
By antisymmetry the derivatives acting on $Q_{\underline{0}}$ are all equivalent up to a sign. Using the $\det$ representation of the VdM we obtain
\begin{align}
{\cal I}_{L,M}\Big|_{(\log\rho)^{ML}}&=   (-)^{ML}\frac{h_{\bf a}(Z,\bar{Z})}{ \prod_{k}\Gamma[L+k]}  \Bigg[ \prod_{i=1}^{M} \partial_i^{q^*_i} 
\det \Big[ (a_i+s_i)^{j-1-L} \Big]_{\ 1\leq i,j\leq M} \Bigg]_{s_i=0}\label{ultimaBD} 
\end{align}
The derivatives and the polynomials $h_{\bf a}=h_{a_1}\ldots h_{a_M}$ can be passed into the determinant. Finally, 
\begin{equation}\label{ultimissimaBD}
{\cal I}_{L,M}\Big|_{(\log\rho)^{ML}}=
(-)^{ML}\det\bigg[  \frac{ ( j-i-L+1)_{i-1}}{\Gamma[L+i]} \times  a_i^{j-i-L} h_{a_i}(Z,\bar{Z}) \bigg]_{\ 1\leq i,j\leq M}
\end{equation}
When $M=1$ we obtain the known result for the ladder
\begin{equation}
{\rm Lad}_{L}\Big|_{(\log\rho)^{L}}:={\cal I}_{L,1}\Big|_{(\log\rho)^{L}}=(-)^L\frac{1}{L!}\frac{h_a(Z,\bar{Z})}{a^L}
\end{equation}
We can then use this projected ladder as a building block in \eqref{ultimissimaBD}, to arrive at
\begin{equation}
{\cal I}_{L,M}=
\det\Big[ (-1)^{i-j} \frac{ ( j-i-L+1)_{i-1}}{(L+i-j+1)_{j-1} } \times  {\rm Lad}_{L+i-j} (Z,\bar{Z})\Big]_{\ 1\leq i,j\leq M}
\end{equation}
where, upon imposing Steinman relations, 
we simply dropped the projection on the maximal log.
Our result looks a bit different from formula presented in \cite{Basso:2017jwq,Basso:2021omx}, but in fact is perfectly equivalent!

\bibliographystyle{apsrev4}

\end{document}